\documentclass[aps,pra,preprint,groupedaddress]{revtex4-2}
\usepackage{graphicx}
\usepackage{dcolumn}
\usepackage{bm}
\usepackage{xcolor,ulem}
\usepackage{float}
\usepackage{bm}
\usepackage{subfigure}
\usepackage{amsmath}
\usepackage{url}
\usepackage[colorlinks, linkcolor=blue,anchorcolor=blue,citecolor=blue,urlcolor=blue,bookmarksnumbered]{hyperref}

\begin{document}
\title{Propagation and excitation properties of nonlinear surface plasmon polaritons in a rectangular barrier}
\author{Xiangchun Tian, Yundong Zhang, Yu Duan, Yong Zhou}
\affiliation{School of Physics and Electronics, Shandong Normal University, Jinan 250014, China}
\author{Chaohua Tan}
\email[E-mail:~]{tanch@sdnu.edu.cn}
\affiliation{School of Physics and Electronics, Shandong Normal University, Jinan 250014, China}
\date{\today}

\begin{abstract}
We propose a scheme to study the nonlinear propagation properties of nonlinear surface plasmon polaritons (SPPs) in a three level $\Lambda$ type electromagnetically induced transparency (EIT) system with modulation of a rectangular barrier. Based on the multi scale method, the nonlinear Schr{\"o}dinger equation (NLSE) describing nonlinear propagation of SPPs is derived, and the rectangular barrier affecting propagation of nonlinear SPPs is provided by an off-resonance Stark field. For the single nonlinear SPPs incident case, by adjusting the height and half width of the barrier, we can realize transmission, trapping and reflection of the nonlinear SPPs. For two nonlinear SPPs symmetrical incident case, we find that a periodic intensity distribution in transverse direction mode can be excited in the rectangular barrier, and we study the relationship between propagation properties of such excited modes in the barrier with nonlinearity, half width of the barrier and phase difference of the initial nonlinear SPPs. In addition, we design an optical switch of nonlinear SPPs based on the above results. The results obtained here not only provide a theoretical basis for the study of the interaction between nonlinear SPPs and external potentials, but also have broad application prospects in the field of optical information at micro/nano scale.
\end{abstract}

\maketitle



\section{Introduction}

Surface plasmon polaritons (SPPs) is a kind of electromagnetic eigenmode which exist in the surface of metal and dielectric material~\cite{RevModPhys.92.025003}. It can confine the electromagnetic fields to the spatial extent which scale is far smaller than the wavelength, therefore, SPPs can overcome the diffraction limitation, and produce strong local field enhancement effect. Recently, SPPs have been widely researched and applied in the fields of light-matter nonlinear interaction enhancement~\cite{verhagen2007enhanced,schuller2010plasmonics,PhysRevB.99.205404,PhysRevLett.120.203903}, optical communication at micro/nano scale~\cite{huang2014electrically,luo2017purcell,choi2017control,zhang2020plasmonic}, nano-focusing~\cite{ropers2007grating,becker2016gap,ZHU2019124358,TAN2021104531}, super resolution display~\cite{zhang2005sub,Takayuki2019white} and so on.

With the further understanding of SPPs, how to manipulate SPPs effectively at micro/nano scale has aroused wide attention. Meanwhile, several kinds of methods have been proposed in order to better modulate the excitation and propagation properties of SPPs:

Firstly, by changing or designing different structures of metals, one can realize manipulation of SPPs ~\cite{PhysRevB.80.035407,Caglayan.08,berini2012surface,Hohenau.05,PhysRevB.77.033417,chen2014ultra,lin2013polarization,PhysRevApplied.9.014032,PhysRevLett.112.023903}, e.g., carving gratings-based structure on the metal surface is an effective way to realize unidirectional control of SPPs~\cite{Caglayan.08,PhysRevB.80.035407}; using the metal film covered with a thin patterned dielectric layer can be used to restrict or redirect SPPs~\cite{berini2012surface,Hohenau.05}; adopting metasurface can also manipulate the SPPs effectively~\cite{PhysRevLett.112.023903,PhysRevApplied.9.014032}.

Secondly, some new materials (such as graphene, sodium) have also been proposed to make dynamic manipulation of SPPs possible~\cite{PhysRevLett.109.073901,PhysRevLett.113.055502,xiong2019photonic,sunku2018photonic,wang2020stable,PhysRevLett.101.263601}. For example, using the negative-index metamaterial (NIMM) instead of metal to excite lower loss surface polaritons (SPs) modes at the dielectric-NIMM interface~\cite{PhysRevLett.101.263601} can realize coherent control of SPs for a long distance.

At last, introducing external potential fields to change the electromagnetic environment in the system can also be used to modulate properties of SPPs~\cite{macdonald2009ultrafast,temnov2010active,PhysRevA.101.023818,PhysRevA.102.063516,melikyan2014high,haffner2015all,Dzedolik.18,Weeber.13,li2019photothermal}. e.g., by using a beam of control light with a specific polarization state to shine on the surface of aluminium, then the refractive index of aluminum will change due to its absorption of control light, thus, the SPPs can be manipulated by the incident light field~\cite{macdonald2009ultrafast}; the phase information and trajectory of SPPs can be modulated by introducing an external gradient magnetic field~\cite{temnov2010active,PhysRevA.101.023818,PhysRevA.102.063516}. Besides, there are some other kinds of external potential fields can be used to realize the modulation of SPPs, such as electric field~\cite{melikyan2014high,haffner2015all,Dzedolik.18} and thermal field~\cite{Weeber.13,li2019photothermal}.

From the above, one can see that the interaction between SPPs and external potential fields has been widely studied experimentally, and some interesting properties of SPPs has been found under the modulation of external potential fields, e.g., the reflection, transmission and trapping of SPPs in the external potentials or interfaces~\cite{PhysRevLett.102.133903,PhysRevA.101.053845}. However, theoretical exploration of interaction mechanism between SPPs and external potentials is relative deficiency, especially for some fundamental cases, such as SPPs with potential wells and SPPs with potential barriers. 

Thus, in this work, we propose a scheme to study interaction between SPPs and a rectangular barrier, including nonlinear propagation and excitation of SPPs in the barrier, in a NIMM-atomic gas interface waveguide system. We first obtain the low loss stable propagation of nonlinear SPPs via electromagnetically induced transparency (EIT).
By introducing a Stark field into the system, a rectangular barrier is produced, we investigate the reflection, tranmission and trapping dynamics of single nonlinear SPPs in the rectangular barrier. Meanwhile, we also obtain the stable propagation modes existed in barrier by inputting two symmetrically incident nonlinear SPPs. Then, we show the influence of several major factors on the propagation modes, such as nonlinearity, width of the barrier and phase difference of initial nonlinear SPPs. At last, we introduce its potential application in optical switch.

The rest of the article is arranged as follows: In Sec.~\ref{Theoretical Model}, we introduce our theoretical model and give the Maxwell-Bloch (MB) equations for the system. In Sec.~\ref{Linear and nonlinear propagation properties of SPPs}, the nonlinear Schr{\"o}dinger equation (NLSE) describing the envelope of the probe field is given. In Sec.~\ref{Propagation properties of SPPs with external potential modulation}, we investigate the propagation properties of SPPs under the manipulation of the rectangular barrier. In final section, we summarize the main results obtained in this work.

\section{Theoretical Model}\label{Theoretical Model}
\begin{figure}[H]
  \centering
  \includegraphics[width=1\textwidth]{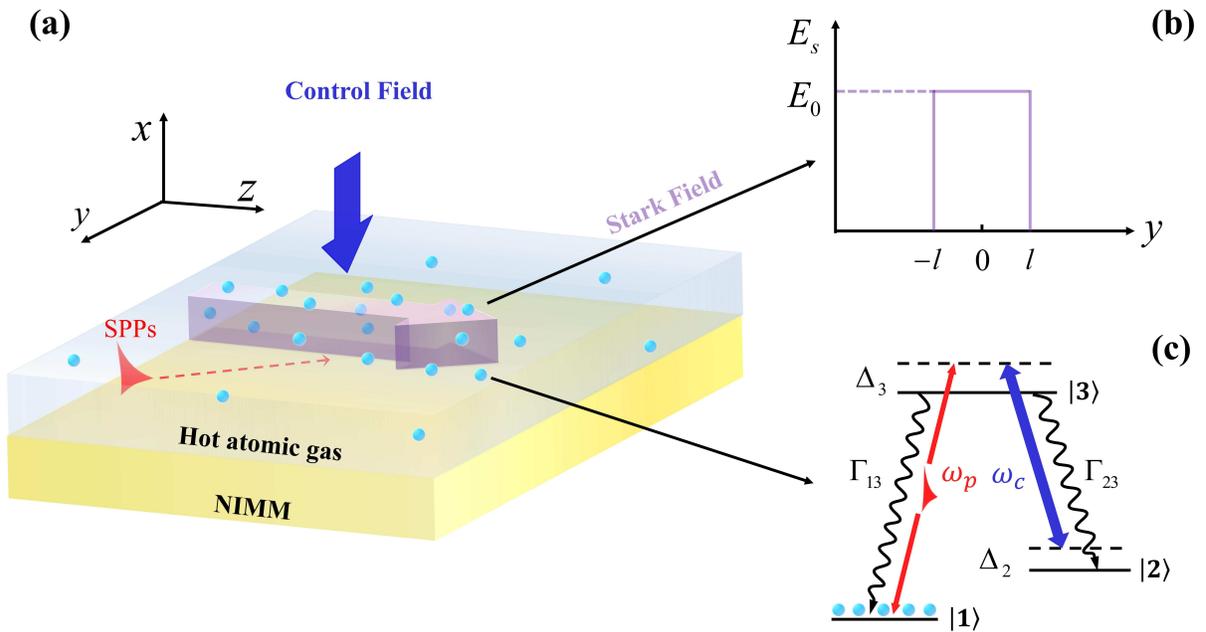}\\
  \caption{\footnotesize (Color online) (a) The schematic of SPPs is excited in the interface between the NIMM and the hot atomic gas and propagates in the $z$ direction. In this waveguide system, the control field (blue arrow) is incident along the vertical direction. The purple arrow stands for the Stark field which distributes as a rectangular function. (b) The distribution function of the Stark field along the y direction. (c) The energy-level diagram of $\Lambda$-type atomic system (the details are given in the text).}\label{fig1}
\end{figure}
The waveguide system we proposed here consists of a layer of NIMM in the lower half plane $x<0$ (with relative permittivity ${\varepsilon _2}$ and relative permeability $\mu _{2}$) and a hot atomic gas filled above the NIMM, as shown in Fig.~\ref{fig1}. The hot atomic gas we using here has a $\Lambda$-type energy level excited configuration, two ground states $\left|1\right\rangle$ and $\left|2\right\rangle$ couple to the excited state $\left|3\right\rangle$ with the weak probe field $\mathbf{E}_p$ with center angle frequency $\omega_p$ (which propagates along $z$ direction and is chosen from guide modes of SPPs at the interface of NIMM) and the strong continuous control field $\mathbf{E}_c$ with center angle frequency $\omega_c$ (which is incident vertically to the waveguide interface), respectively. ${\Delta_3}$ (${\Delta_2}$) is the one (two) photon detuning. The atoms occupy on the state $\left|3\right\rangle$ can spontaneously radiate to the ground states with the rate $\Gamma_{j3}$ ($j=1,2$). The two ground states are chosen from the hyperfine structure, thus, the transition between the two ground states is forbidden.

The NIMM interface waveguide system can support both transverse electric (TE) and transverse magnetic (TM) modes, the TE modes decay rapidly in the system, however, the TM modes can propagate without loss under specific conditions~\cite{PhysRevLett.101.263601}.
Thus, we choose the lowest order TM mode as the probe field of our system, which can be expressed as $\mathbf{E}_p( \mathbf{r},t) = {{\mathcal E}_p}(y,z,t)\mathbf{u}_p(x){e^{i\left[{{k (\omega _p)}z-{\omega_p}t}\right]}}+c.c.$, where ${{\mathcal E}_p}$ is the envelope of the probe field, $\mathbf{u}_{p}(x)$ is the mode function in the $x$ direction (which is given in Appendix~\ref{Appendix A}), ${k}(\omega _p) = \left( {{\omega _p}/c} \right){\left[ {{\varepsilon _1}{\varepsilon _2}\left( {{\varepsilon _1}{\mu _2} - {\varepsilon _2}{\mu _1}} \right)/\left( {\varepsilon _1^2 - \varepsilon _2^2} \right)} \right]^{1/2}}$ is the propagate constant of the probe field without modulation of the atomic gas. The control field reads ${\mathbf{E}_c}\left( {\mathbf{r},t} \right) =   {{\mathcal E}_c}{\mathbf{e}_c}{e^{i\left( {{k_c}x - {\omega _c}t} \right)}}+c.c.$, in which ${{\mathcal E}_c}$ is strength, $\mathbf{e}_{c} $ is the unit polarization vector and $k_c$ is the propagate constant of the control field.

Meanwhile, we apply another off-resonance laser field (Stark field) propagating along the interface of the NIMM in $z$ direction to generate a rectangular barrier in $y$ direction, with center frequency ${\omega _s}$, and reading
\begin{equation}
  {\mathbf{E}_s}(y,t) = {\mathbf{e}_s}\sqrt 2 {E_s}(y)\cos ({\omega _s}t),
\end{equation}
where ${E_s}(y)$ is the amplitude of the Stark field and $\mathbf{e}_{s} $ is the unit polarization vector of the Stark field. Such a Stark field can produce a small energy level shift to state $\left| j \right\rangle $ in $y$ direction, as $\Delta {E_j} =  - {\alpha _j}{\left\langle {\mathbf{E}_s^2} \right\rangle _t}/2=-{\alpha _j}{\left| {{E_s}\left( y \right)} \right|^2}/2$, in which $\alpha_{j}$ is the scalar polarizability of the state $\left| j \right\rangle $, ${\left\langle {\mathbf{E}_s^2} \right\rangle _t}=\int_0^T {\mathbf{E}_s^2} dt / T $ is the time average in an oscillating cycle. After taking the energy level shift caused by the Stark field into account, the optical detunings should be corrected to ${\Delta _j}^\prime  = {\Delta _j} + {\alpha _{j1}}{\left| {{E_s}\left( y \right)} \right|^2}/2$, with ${\alpha _{jl}} = \left( {{\alpha _j} - {\alpha _l}} \right)/\hbar $.

In the interaction picture, under electric-dipole and rotating-wave approximation, the Hamiltonian of the system is given by
\begin{equation}
{\hat H_{{\rm{int}}}} =  - \hbar \left[ {\sum\limits_{j = 1}^3 {{\Delta _j}^\prime \left| j \right\rangle \langle j|}  + {\Omega _c}\left| 3 \right\rangle \langle 2| + {\zeta}\left( x \right) e^{i\theta _p} {\Omega _p}\left| 3 \right\rangle \langle 1| + \rm H.\rm c.} \right],
\end{equation}
where ${\Omega _p} = \left| {{{\mathbf{p}}_{13}}} \right|{\mathcal{E}_p}/\hbar $, ${\Omega _c} = \left| {{{\mathbf{p}}_{23}}} \right|{\mathcal{E}_c}/\hbar $, $ \zeta \left( x \right) = {\mathbf{e}_{13}} \cdot {\mathbf{u}_p}\left( x \right)$, and $\mathbf{p}_{ij}$ is the electric-dipole matrix element associated with the state $\left| i \right\rangle $ to state $\left| j \right\rangle $, $\theta _p=[k(\omega_p) + k_1 - k_3] z$ is the phase mismatch caused by the dispersion of SPPs, $k_j$ ($j$=1, 2, and 3) are the wave number of state $\left| j \right\rangle$. The motion of atoms is governed by the optical Bloch equation
\begin{equation}\label{Bloch equation}
  \frac{{\partial \sigma }}{{\partial t}} =  - \frac{i}{\hbar }\left[ {{{\hat H}_{{\mathop{\rm int}} }},\sigma } \right] - \Gamma \sigma,
\end{equation}
where $\sigma$ is a $3\times3$ density matrix, $\Gamma $ is a $3\times3$ relaxation matrix describing the spontaneous emission and dephasing effect of the system. The explicit expressions of Eq.~(\ref{Bloch equation}) are given in Appendix~\ref{Appendix B}.

The dynamics evolution of the probe field in this system is governed by the Maxwell equations. Under the slowly varying envelope approximation, the Maxwell equation can be reduced to
\begin{equation}\label{Maxwell equation}
i\left( {\frac{\partial }{{\partial z}} + \frac{1}{c}\frac{1}{{{n_{\rm eff}}}}\frac{\partial }{{\partial t}}} \right){\Omega _p} {e^{i\theta _p}} + \frac{c}{{2{\omega _p}}}\frac{1}{{{n_{\rm eff}}}}\frac{{{\partial ^2}}}{{\partial {y^2}}}{\Omega _p} {e^{i\theta _p}} + {\kappa _{13}}\int_{ - \infty }^\infty  {d(kv)f\left( kv \right)\left\langle {{\sigma _{31}}} \right\rangle }  = 0 ,
\end{equation}
where ${n_{\rm eff}} = c{k(\omega_p)}/{\omega _p}$ is the effective refractive index, ${\kappa _{13}} = {{{\mathcal{N}_\alpha }{{\left| {{\mathbf{p}_{13}}} \right|}^2}\omega _p^2} \mathord{\left/ {\vphantom {{{\mathcal{N}_\alpha }{{\left| {{\mathbf{p}_{13}}} \right|}^2}\omega _p^2} {[2{\varepsilon _0}{c^2}\hbar \tilde k({\omega _p})]}}} \right.\kern-\nulldelimiterspace} {[2{\varepsilon _0}{c^2}\hbar \tilde k({\omega _p})]}}$ is the coupling coefficient describing the interaction between the weak probe field and the atoms, ${\mathcal{N}_\alpha }$ represents the number density of the atoms and $\tilde k = {\mathop{\rm Re}\nolimits} \left( k \right)$. The expectation operator $\left\langle {...} \right\rangle $ is defined as $\left\langle {\psi \left( x \right)} \right\rangle  \equiv \int_{ - \infty }^\infty  {dx{\zeta ^*}\left( x \right)\psi \left( x \right)} /\int_{ - \infty }^\infty  {dx{{\left| {\zeta \left( x \right)} \right|}^2}} $, i.e., average over mode distribution function in $x$ direction.
In addition, the hot atomic gas is above the NIMM, thus, the Doppler effect must be taken into account, which could cause an inhomogeneous broadening of the energy level. In Eq.~(\ref{Maxwell equation}), we have averaged the inhomogeneous broadening due to Doppler effect over the Maxwell velocity distribution $f(kv)$~\cite{Figueroa.06}, and $f(kv)$ reads
\begin{equation}
  f\left( kv \right) = \frac{{\sqrt {\ln 2} }}{{\sqrt \pi  {W_D}}}{e^{ - \ln 2{{\left( {\frac{{kv}}{{{W_D}}}} \right)}^2}}} ,
\end{equation}
where ${W_D} = k\sqrt {2\ln 2{k_B}T/m} $ is the full width at the half-maximum (FWHM) of the inhomogeneous broadening, $k_{B}$ is the Boltzmann constant, $T$ and $m$ are the temperature and mass of the atom, respectively.

Equations~(\ref{Bloch equation}) and~(\ref{Maxwell equation}) are known as MB equations which describe the interaction dynamics of the system.

\section{Derivation of the Nonlinear Schr{\"o}dinger Equation}\label{Linear and nonlinear propagation properties of SPPs}

In EIT system, the probe field is weaker than the control field, we can adopt the standard multiple scales method~\cite{PhysRevE.72.016617} to solve the MB equations.  The asymptotic expansion are given as $\sigma_{jl}-\sigma_{jl}^{(0)}=\epsilon\sigma_{jl}^{(1)}+\epsilon^2\sigma_{jl}^{(2)}+\epsilon^3\sigma_{jl}^{(3)}+...$, $\Omega_p= \epsilon\Omega_p^{(1)}+\epsilon^2\Omega_p^{(2)}+...$,
where, $\epsilon$ is a dimensionless small parameter characterizing the typical amplitude ratio of the probe and the control field, and the physical quantities in the left of the expressions can be treated as the functions of the multiple scales variables, with $z_j=\epsilon^jz$ $(j=0,2)$, $y_j=\epsilon^j y$ $(j=1)$ and $t_j=\epsilon^jt$ $(j=0,2)$. In our analysis, we assume that the amplitude of Stark field can be written as $E_s=\epsilon E_s^{\left( 1 \right)}$. Then $d_{ij}$ can be expressed as $d_{ij} = d_{ij}^{\left( 0 \right)} + {\epsilon ^2}d_{ij}^{\left( 2 \right)}$, with $d_{ij}^{\left( 2 \right)} = {\alpha _{ij}}{\left| {E_s^{(1)}} \right|^2}/2$. Substituting the above expansion into the MB equation, we can obtain a series linear but inhomogeneous equations order by order.

The zeroth order solution corresponds to the steady state of the system, i.e., the probe field is absent. The specific expression of the zeroth order solution reads $\sigma ^{(0)}_{11}=1$, and the zeroth order of all the other elements of the density matrix are equal to zero, i.e., $\sigma_{ij}^{(0)}=0$, which means that, when the system is prepared, due to the pumping effect of the strong control field, all the atoms occupy on the state $|1\rangle$, and there are no coherence between all the states.

In the first order solution, we can obtain the linear propagation property of the probe field. When the probe field is incident, the population and coherence in the system should be modified. However, due to $\Omega_p$ is taken as a small quantity, thus, the population on state $|1\rangle$ is not changed.
We assume that the solution of first order of the probe field has the following form $\Omega_p^{(1)}=Fe^{i\theta}$, where $\theta=K(\omega)z_0-\omega t_0$, $F$ is the slowly varying envelop function of multiscale variables, $K$ is the linear dispersion relation of the probe field, the imaginary part of $K$ means linear absorption of the system and the real part of $K$ means linear dispersion of the system. The expression of the linear dispersion relation $K(\omega)$ reads
\begin{equation}\label{K}
{K}\left( \omega  \right) = \frac{\omega }{c}\frac{1}{{{n_{\rm eff}}}} + {\kappa _{13}}\int_{ - \infty }^\infty  {d(kv)f\left( kv \right)\left\langle {\frac{{\left( {\omega  + d_{21}^{\left( 0 \right)}} \right)}}{{{D}}}\zeta \left( x \right)} \right\rangle },
\end{equation}
where ${{D} = {{\left| {{\Omega _c}} \right|}^2} - \left( {\omega  + d_{21}^{\left( 0 \right)}} \right)\left( {\omega  + d_{31}^{\left( 0 \right)}} \right)}$, $\omega$ is the frequency shift to the center frequency of the probe field $\omega_p$, and $K(\omega)$ actually is also the wavenumber shift to $k(\omega_p)$.

The absorption spectrum Im$(K)$ of the probe field will open a transparency window at the center frequency for a large control field, i.e., EIT effect. Although the EIT effect can effectively inhibit the linear absorption of the medium to the probe field and slow down the group velocity (which is defined as ${V_{g}} = {\left[ {{{\partial {\rm Re}({K}\left( \omega  \right))} \mathord{\left/{\vphantom {{\partial {K}\left( \omega  \right)} {\partial \omega }}} \right. \kern-\nulldelimiterspace} {\partial \omega }}} \right]^{ - 1}}$) of the probe field, the diffraction and nonlinear effects will lead to serious deformation of the envelope of the probe field in the propagation process. Therefore, we solve the MB equations up to third order to investigate the nonlinear propagation and interaction of the probe field in the rectangular potential produced by the Stark field.

In the third-order approximation equations, we obtain the solvable condition of $F$, which reads
\begin{equation}\label{Nonlinear Schordinger equation}
i\left( {\frac{\partial }{{\partial {z_2}}} + \frac{1}{{{V_{g}}}}\frac{\partial }{{\partial {t_2}}}} \right){F} + \frac{c}{{2{\omega _p}}}\frac{1}{{{n_{\rm eff}}}}\frac{{{\partial ^2}}}{{\partial y_1^2}}{F} + {W_1}{\left| {{F}} \right|^2}{F}{e^{ - 2{{\tilde \alpha }}{z_2}}} + {W_2}{\left| {E_s^{\left( 1 \right)}\left( y \right)} \right|^2}{F} = 0  ,
\end{equation}
with ${{\tilde \alpha }} = {\epsilon ^{ - 2}} \alpha= {\epsilon ^{ - 2}}{\mathop{\rm Im}\nolimits} \left[ {{K}\left( \omega  \right)} {+ k(\omega _p)} \right]$, $W_1$ is the nonlinear coefficient describing the self-phase modulation of the probe field and $W_2$ is the external potential modulation coefficient, specific expressions are given in Appendix~\ref{Appendix C}. By defining  $U=\epsilon F e^{-\alpha z}$ and returning the variables in Eq.~(\ref{Nonlinear Schordinger equation}) to their original scale, one can obtain the equation describing evolution of envelope $F$ of the probe field, which reads
\begin{equation}\label{nls}
i\left( {\frac{\partial }{{\partial z}} + \frac{1}{{{V_g}}}\frac{\partial }{{\partial t}}} \right)U + \frac{c}{{2{\omega _p}}}\frac{1}{{{n_{\rm eff}}}}\frac{{{\partial ^2}}}{{\partial {y^2}}}U + {W_1}{\left| U \right|^2}U + {W_2}{\left| {{E_s}\left( y \right)} \right|^2}U =  - i\alpha U .
\end{equation}

Generally, the coefficients of Eq.~(\ref{nls}) are complex, and $\alpha\neq 0$, which means the linear absorption and nonlinear dissipation in our system will hugely affect the propagation of the probe field. In order to study excitation and propagation of nonlinear SPPs in the rectangular barrier, the probe field need to propagate for a long distance stably. Thus, firstly, the linear and nonlinear absorption of the system must be suppressed, secondly, the diffraction effect of the probe field must be balanced by the nonlinearity provided by the system.

However, when the system works under EIT condition, a realistic set of physical parameters, which will be shown in the following, can be found to make the imaginary part of the coefficients in Eq.~(\ref{nls}) to be negligible small compare to its corresponding real parts, together with $\alpha \approx0$, and a giant Kerr nonlinearity.

Based on the above condition, we define a set of dimensionless variables as $\xi  = \sqrt {{n_{\rm eff}}} y/{R_y}$, $s = z/{L_D}$, $\tau  = t/{\tau _0}$, $u = U/{u_0}$, $g = {V_{g}}{\tau _0}/{L_D}$, ${w_1} = {L_D}/{L_N}$, ${w_2} = {L_D} {\rm Re} ({W_2})E_0^2$, $V\left( \xi  \right) = {\left( {{E_s}/{E_0}} \right)^2}$, with ${L_D} = {\omega _p}R_y^2/c$ being the typical diffraction length, $R_y$ being the transverse width of the SPPs in the $y$ direction, and $L_N=1/\left[ {{\mathop{\rm Re}\nolimits} \left( {{W_1}} \right)u_0^2} \right]$ being the typical nonlinear length, $E_0$ being a real parameter representing typical strength of the Stark field. Then, Eq.~(\ref{Nonlinear Schordinger equation}) can be simplified as the form of the dimensionless NLSE
\begin{equation}\label{Dimensionless nonlinear Schordinger equation}
  i\left( {\frac{\partial }{{\partial s}} + \frac{1}{g}\frac{\partial }{{\partial \tau }}} \right)u + \frac{1}{2}\frac{{{\partial ^2}u}}{{\partial {\xi ^2}}} + {w_1}{\left| u \right|^2}u + {w_2}V\left( \xi  \right)u =  0 .
\end{equation}
To solve the above equation, we assume the solution  as $u\left( {\tau ,s,\xi } \right) = \lambda \left( {\tau ,s} \right)v\left( {s ,\xi } \right)$ based on separated variable method. For $\lambda \left( {\tau ,s} \right)$, we take it as a Gaussian profile $\lambda \left( {\tau ,s} \right) = {\left[ {1/({\rho _0}\sqrt \pi  )} \right]^{1/2}}\exp \left[ { - \rho^2/\left( {2\rho _0^2} \right)} \right]$, with $\rho=\tau - s/g$, and ${\rho _0}$ is a free real parameter~\cite{PhysRevA.78.013833,PhysRevA.86.043809}. Then, Eq.~(\ref{Dimensionless nonlinear Schordinger equation}) can be further simplified as
\begin{equation}\label{NLSE2}
  \left( {i \frac{\partial }{{\partial s }} + \frac{1}{2}\frac{{{\partial ^2}}}{{\partial {\xi ^2}}}} \right)v + {w_1}{\left| v \right|^2}v + {w_2}V\left( \xi  \right)v =  0.
\end{equation}
It has been proved that, Eq.~(\ref{NLSE2}) can be used to describe the propagation and interaction of nonlinear optical pluse with the modulation of the external potential.

\section{Propagation properties of nonlinear SPPs interacting with the rectangular barrier}\label{Propagation properties of SPPs with external potential modulation}

Next, we choose a realistic system to carry out our study. The NIMM is selected as a sliver based material, the relative permittivity and permeability of the NIMM can be described by the Drude model in optical region, i.e., ${\varepsilon _2}\left( \omega  \right) = {\varepsilon _\infty } - \omega _e^2/\left[ {\omega \left( {\omega  + i{\gamma _e}} \right)} \right]$,
${\mu _2}\left( \omega  \right) = {\mu _\infty } - \omega _m^2/\left[ {\omega \left( {\omega  + i{\gamma _m}} \right)} \right]$,
where ${\omega _e}$ ($\omega _m$) is the electric (magnetic) plasma frequency, ${\gamma _e}$ ($\gamma _m$) is the electric (magnetic) decay rate, ${\varepsilon _\infty }$ and ${\mu _\infty }$ are the background constant. The related parameters are ${\varepsilon _1} = 1,{\mu _1} =  1$, ${\varepsilon _\infty } = {\mu _\infty } = 1$,  $\omega_{e}/2\pi=2.18\times10^{15} ~ \rm{s^{-1}}$, $\gamma_{e}/2\pi=4.35\times10^{12} ~ \rm{s^{-1}}$, $\omega_{m}=\omega_{e} / 5$, $\gamma _{m} = \gamma _{e} / 1000 $ (as for Ag)~\cite{PhysRevLett.101.263601}. The $\Lambda$ type energy level system is chosen from D2-line of $^{87} \rm{Rb}$ atomic gas at room temperature, with the levels $\left| 1 \right\rangle  = \left| {5{}^2{S_{1/2}},~F = 1,~{m_F} = 0} \right\rangle $, $\left| 2 \right\rangle  = \left| {5{}^2{S_{1/2}},~F = 2,~{m_F} = 0} \right\rangle $ and $\left| 3 \right\rangle  = \left| {5{}^2{P_{3/2}},~F = 2,~{m_F} = 1} \right\rangle $~\cite{steck2001rubidium}. Thus, ${\lambda _c} \simeq {\lambda _p} = 780 ~ \rm{nm}$, $\left| {{\mathbf{p}_{13}}} \right|  \simeq \left| {{\mathbf{p}_{23}}} \right|\simeq 1.27 \times {10^{ - 27}} ~ \rm{C \cdot cm}$, ${\Gamma _3} \simeq 2\pi  \times 6 ~ \rm{MHz}$, $\Gamma_{13}=\Gamma_{23}=0.5\Gamma_3$, and the Doppler width is $2{W_D} = 3.21~\rm{GHz}$. We assume that the atomic density is ${\mathcal{N}_\alpha }=7.18 \times {10^{13}} ~ \rm{c{m^{ - 3}}}$ and obtain $\kappa_{13}=5 \times {10^{11}} ~ \rm{c{m^{ - 1}}{s^{ - 1}}}$.

In this case, the wavenumber of the probe field without modulation of the atomic gas $k(\omega_p)=(8.05\times10^{4}+0.053i)$ cm$^{-1}$. We choose the control field ${\Omega _c} = 2\pi  \times 0.2~\rm{GHz}$, the detunings ${\Delta _2} =  - 2\pi  \times 40.25 ~\rm{MHz}$, ${\Delta _3} =  - 2\pi  \times 31.11 ~\rm{MHz}$. Then, the coefficients of Eq.~(\ref{nls}) are ${W_1} = \left( {7.53 - 0.89i} \right) \times {10^{ - 17}} ~\rm{c{m^{ - 1}}{s^2}}$, ${W_2} = \left( {1.26 - 0.02i} \right) \times {10^{ - 9}} ~\rm{c{m^{ - 1}}{V^2}}$ and the group velocity is ${V_{g}} = 7.69 \times {10^{ - 5}}c$. The total linear absorption of the probe field is $\alpha = {\rm {Im}} [K(\omega) + k(\omega_p)]=0.37$ cm$^{-1}$, thus, the typical absorption length $L_A=1/ \alpha =2.69$ cm. In addition, by taking $\tau_0=0.25 \times {10^{ - 7}} ~\rm{s}$ and $R_y=8.46 ~\rm{\mu m}$, one can obtain the typical diffraction length and the nonlinear length ${L_D} \simeq {L_N} = 0.058 ~\rm{cm}$, and $w_1 = w_2 = 1$. It is easy to see that the imaginary parts of these coefficients in Eq.~(\ref{nls}) are negligible small compared to their corresponding real parts, thus we can ignore the effect of these imaginary parts, together with linear absorption of the system. The following discussion of SPPs propagation and excitation in a rectangular barrier is based on the above set of parameters.

As we mentioned above, the influence of the Stark field is reflected as the external potential term $V\left( \xi  \right)$ in Eq.~(\ref{NLSE2}), which can produce a spatial modulation to the propagation of nonlinear SPPs. In the following, we will investigate the propagation properties of nonlinear SPPs interacting with the external potential based on the Eq.~(\ref{NLSE2}). A intensity uniform Stark field with a transversely finite width can produce a $V(\xi)$ with the form as a rectangular function (a finite high barrier), which reads
\begin{equation}
V\left( \xi  \right) = \left\{ {\begin{array}{*{20}{c}}
{{V_0}}&{{\rm{for}}}&{\left| \xi  \right| \le l}\\
0&{{\rm{for}}}&{\left| \xi  \right| > l}.
\end{array}} \right.
\end{equation}
where $V_0$ and $l$ are the height and half width of the potential barrier $V(\xi)$, respectively.

In the case of $w_1 >0$ and $V(\xi)=0$, Eq.~(\ref{NLSE2}) supports bright soliton solutions with the form as
\begin{equation}\label{bright soliton solution}
v = {\varsigma _0}{\mathop{\rm sech}\nolimits} \left[ {{\varsigma _0}\left( {\xi  - {\eta _0} s  - {\xi _0}} \right)} \right]{e^{i\left[ {{\eta _0}\xi  - \left( {\eta _0^2 - \varsigma _0^2} \right) s /2 - {\varphi _0}} \right]}},
\end{equation}
where $\varsigma _0,\eta_0 ,{\xi _{0}}$ and $\varphi _{0}$ are the free parameters~\cite{PhysRevA.100.013827}. Thus, we take this solution as the initial condition of the probe field to investigate the propagation and excitation properties of SPPs interacting with the rectangular barrier.

\subsection{The transmission, trapping and reflection of a nonlinear SPPs}

\begin{figure}[H]
\centering
\includegraphics[height=0.4\textwidth, width=0.45\textwidth]{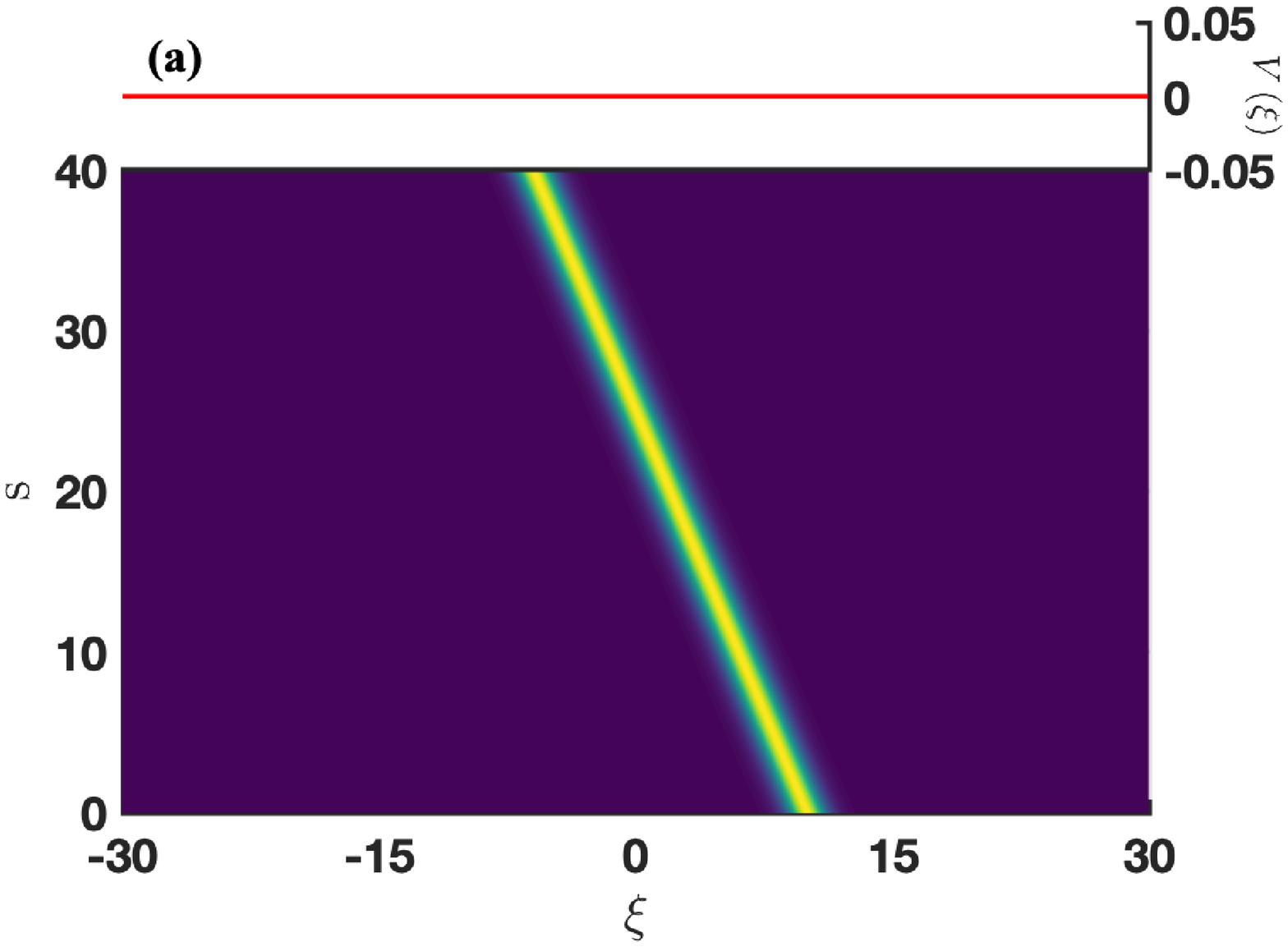}
\includegraphics[height=0.4\textwidth, width=0.45\textwidth]{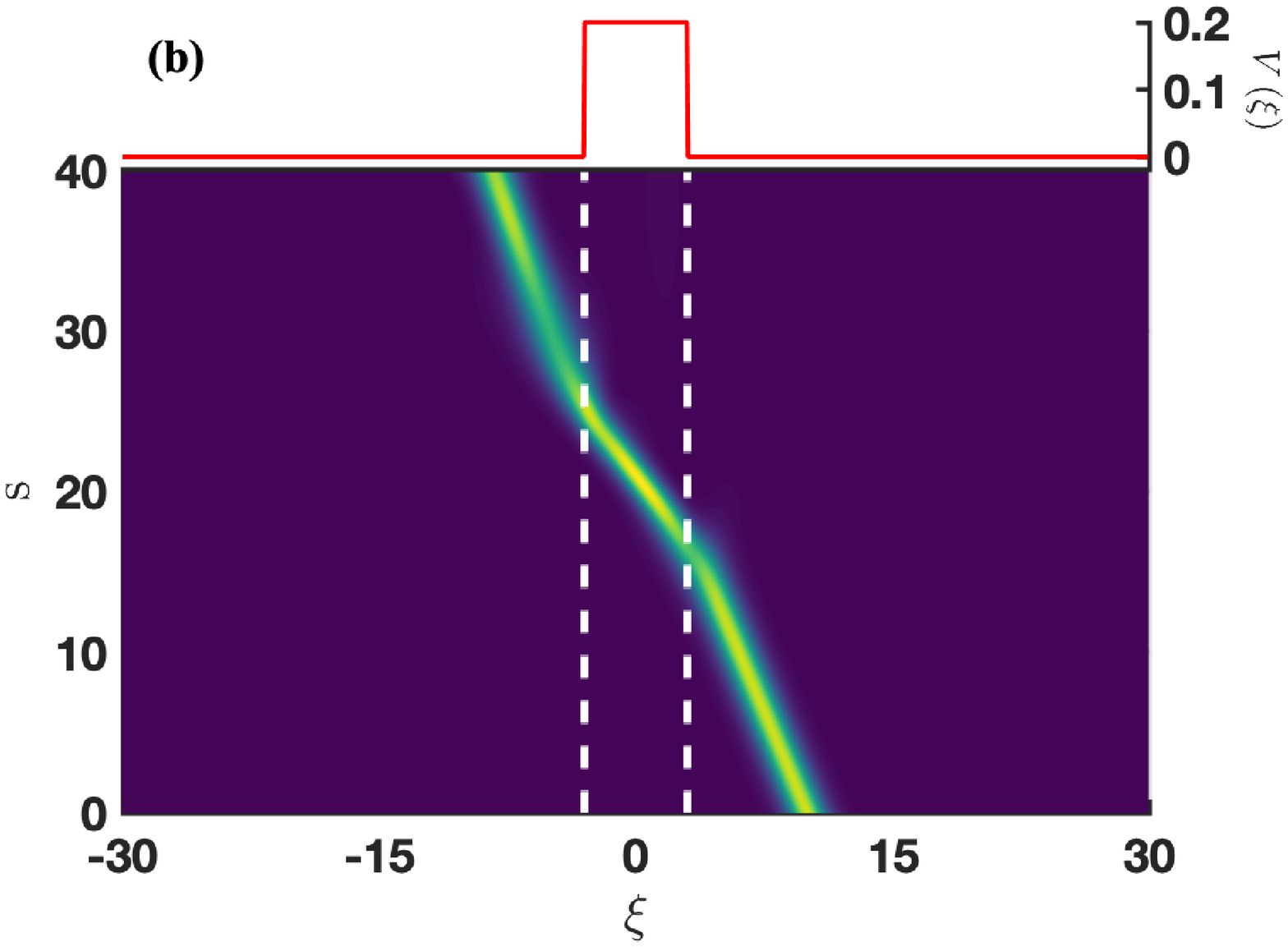}
\includegraphics[height=0.4\textwidth, width=0.45\textwidth]{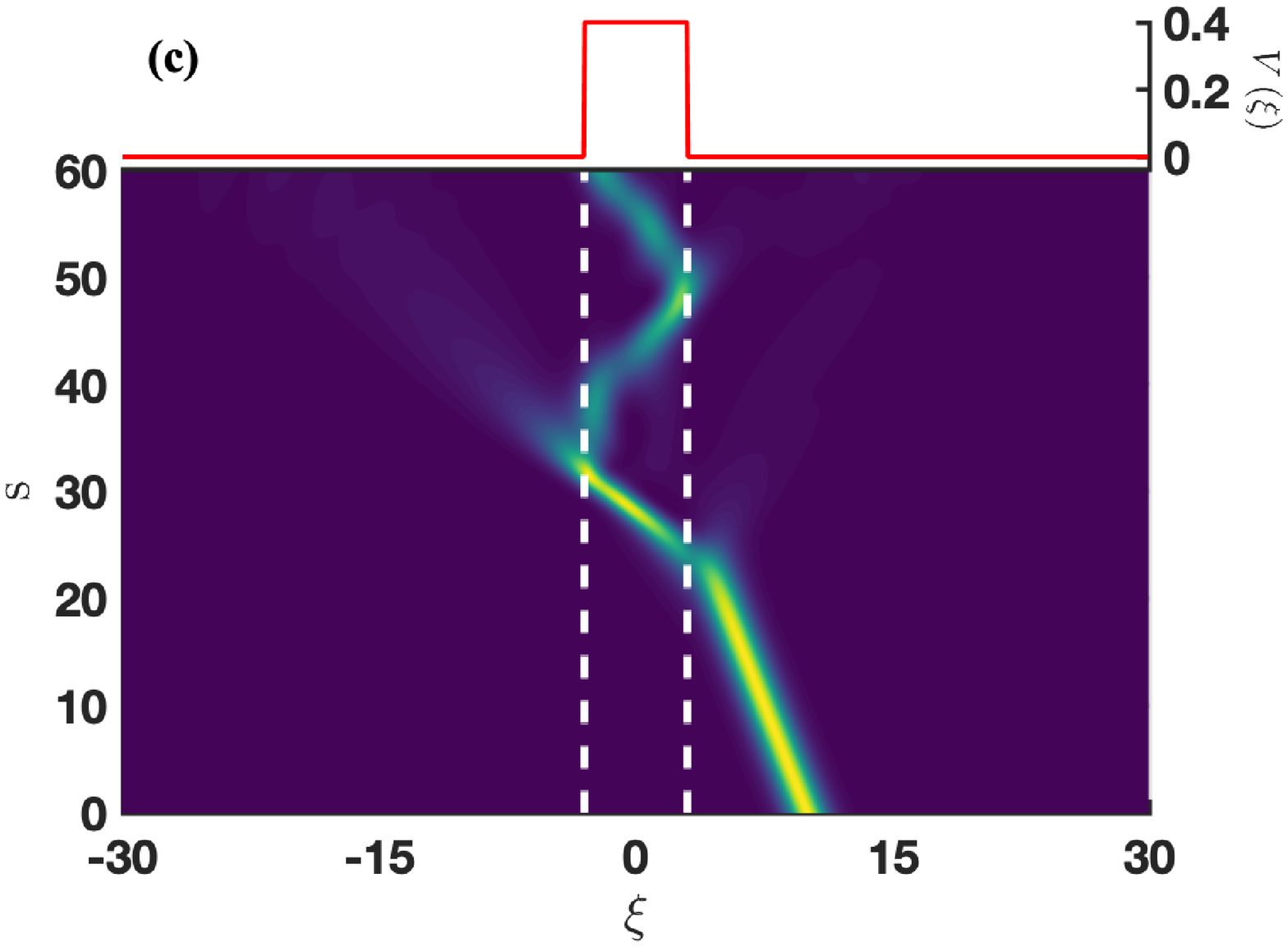}
\includegraphics[height=0.4\textwidth, width=0.45\textwidth]{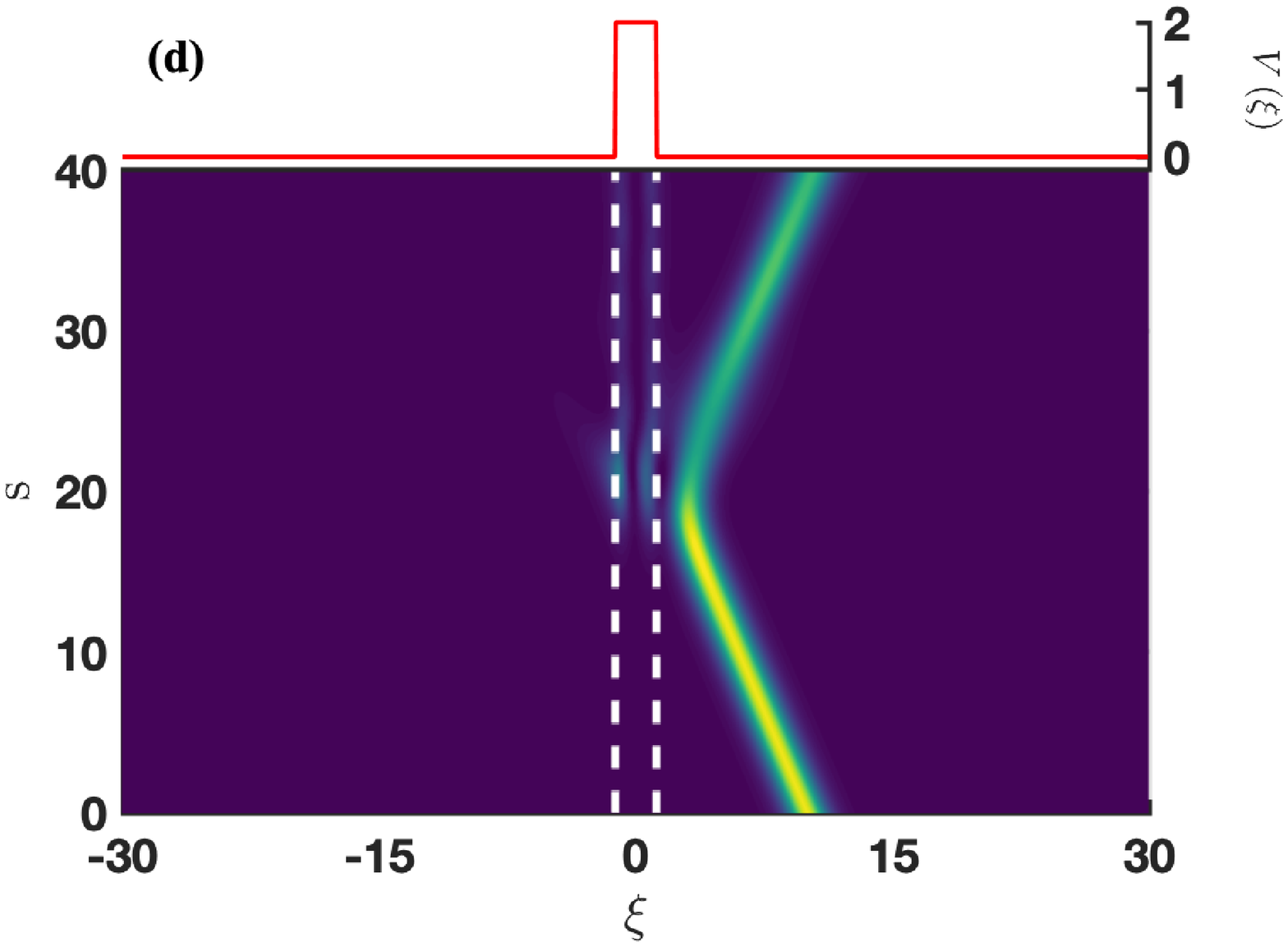}
\caption{The propagation properties of single nonlinear SPPs in the rectangular barrier. (a) The stable evolution process of a nonlinear SPPs without $V(\xi)$. (b) The tranmission process of a nonlinear SPPs through the rectangular barrier. (c) The trapping process of a nonlinear SPPs in the rectangular barrier. (d) The reflection process of a nonlinear SPPs at the right boundary of the rectangular barrier.}
\label{fig2}
\end{figure}

Shown in Fig.~\ref{fig2} are simulation results of inputting the nonlinear SPPs with the form of single bright soliton. In Fig.~\ref{fig2}(a), the specific initial condition of the nonlinear SPPs is $v={\rm sech}(\xi-10) e^{-i0.4 \xi}$, we can see that, when the rectangular barrier is absent, i.e., $V(\xi)=0$, as red line in inset of Fig.~\ref{fig2}(a), the nonlinear SPPs can propagate stably for a long distance due to the balance between the diffraction effect and the nonlinearity of the SPPs.

Shown in Fig.~\ref{fig2}(b) is the transmission of the nonlinear SPPs through the rectangular barrier. The specific initial condition of the nonlinear SPPs is the same as Fig.~\ref{fig2}(a). The height and half width of the rectangular barrier are $V_0=0.2$ and $l=3$, respectively, the profile of $V(\xi)$ is as the red line in the top of Fig.~\ref{fig2}(b), and the white dashed lines denote the boundary of the barrier. The nonlinear SPPs first contacts the right boundary, and partial energy of the SPPs is reflected out of the system, but we can see that the energy loss is negligible, the nonlinear SPPs can keep its shape and pass through the barrier with a phase shift.

However, when enhancing the strength of the rectangular barrier and decrease the incident angle of the nonlinear SPPs, as shown in Fig.~\ref{fig2}(c), $V_0=0.4$ and $v={\rm sech}(\xi-10) e^{-i0.25 \xi}$. The nonlinear SPPs can also go though the right boundary of the barrier and its most energy can go into the barrier, then the SPPs keep going, and on the left boundary, due to decay of amplitude of the SPPs, it is mainly reflected, and the pulse coming out from the left boundary part quickly broadens due to the diffraction effect. Inside the barrier, the SPPs is reflected back and forth on the left and right boundaries with continuous attenuation of amplitude, thus, we can realize the trapping of SPPs, but the effective propagation length of SPPs is short due to the energy loss on the boundary. In Fig.~\ref{fig2}(d), the height and half width of the rectangular barrier are $V_0=2$ and $l=1.2$, respectively. For such a high and narrow barrier, we choose a specific initial condition which is the same as Fig.~\ref{fig2}(b), we can see that the nonlinear SPPs is mainly reflected on the right boundary. Therefore, by adjusting the height and half width of the rectangular barrier, we can realize the transmission, trapping and reflection of the nonlinear SPPs.

\subsection{Excitation of stable propagation modes of nonlinear SPPs in the rectangular barrier}

\begin{figure}
\includegraphics[width=0.45\textwidth]{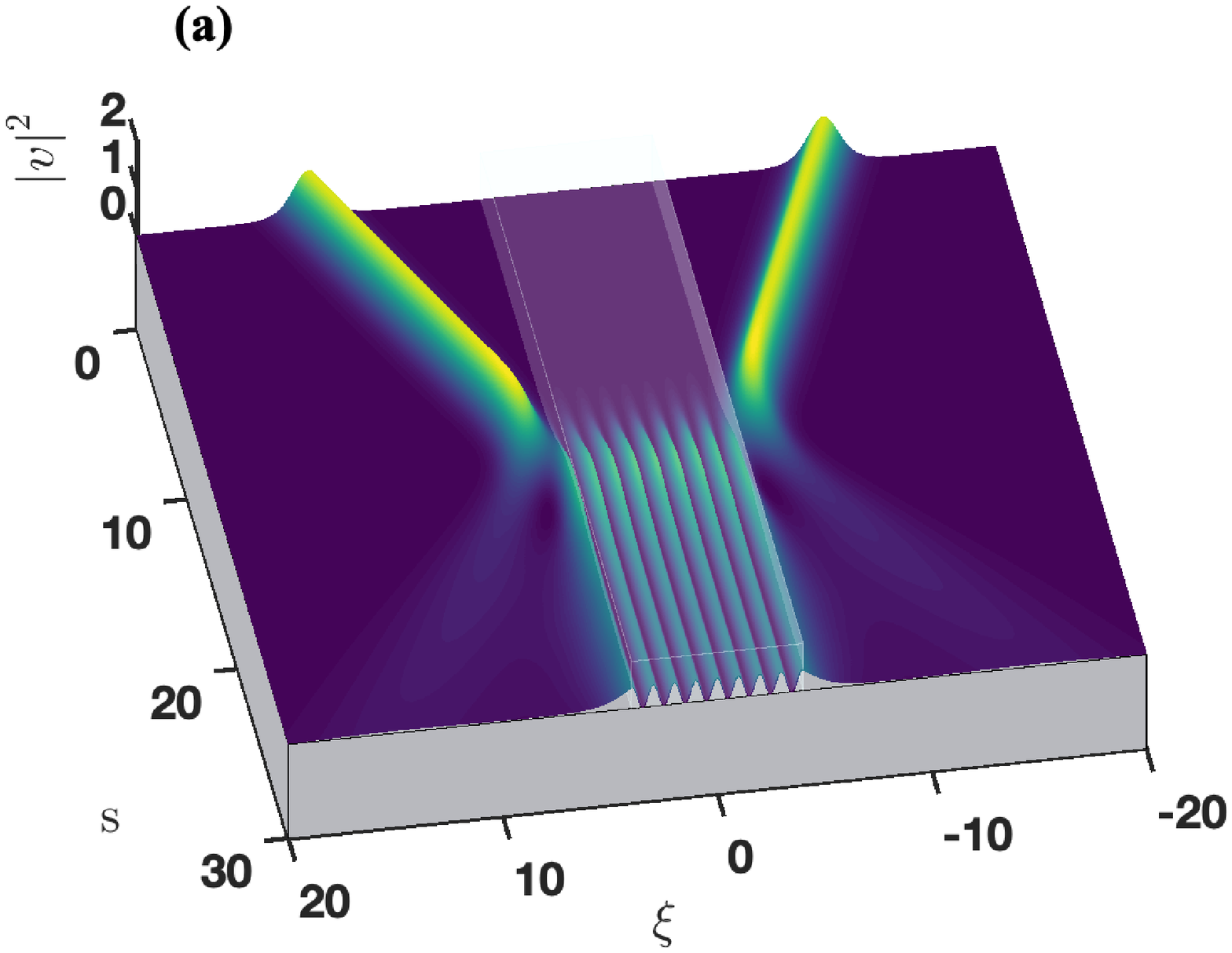}
\includegraphics[width=0.45\textwidth]{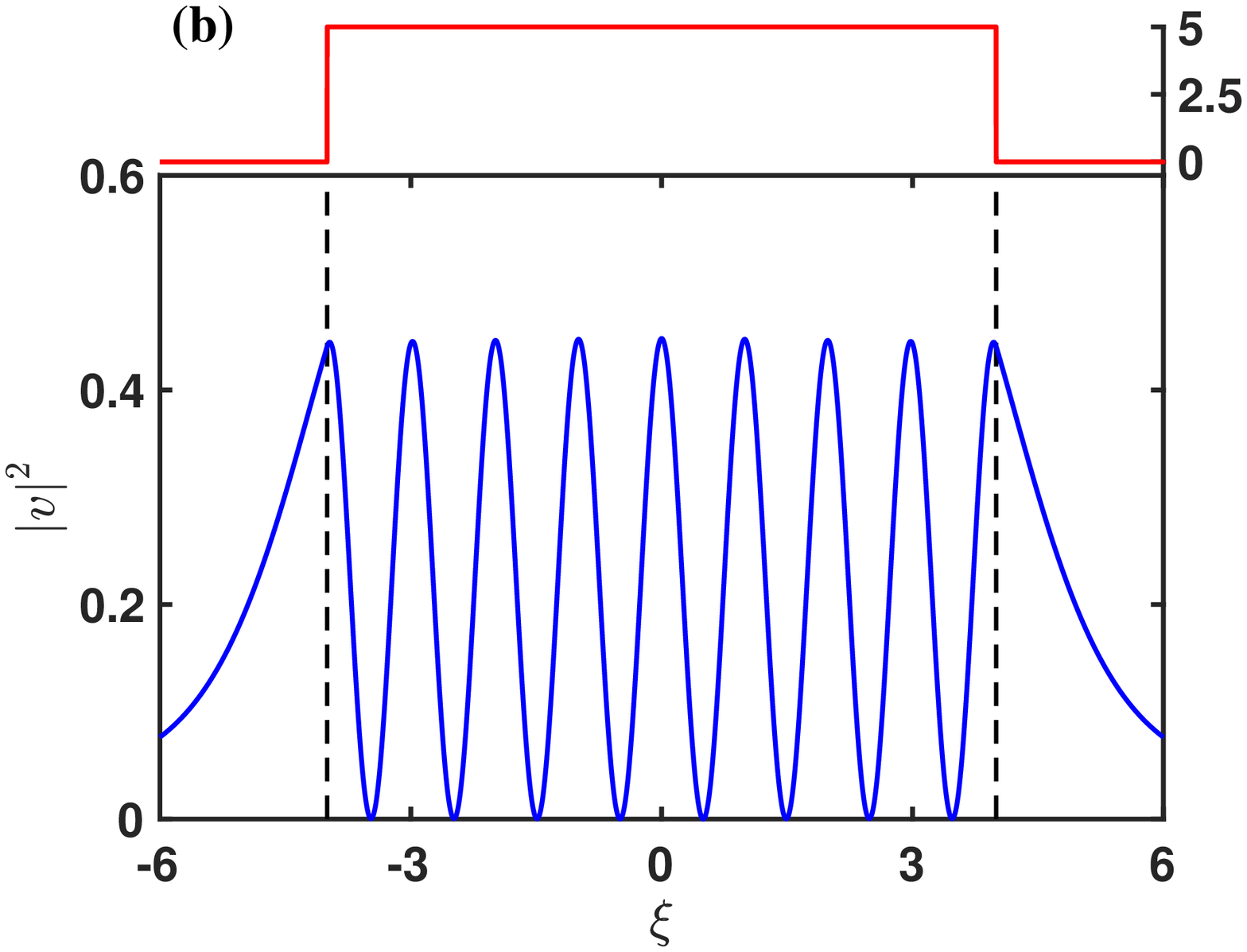}
\caption{(a) The periodic propagation modes which are excited by two symmetrically incident SPPs in the rectangular barrier. (b) The profile of intensity $|v|^2$ as a function of $\xi$ at $s=30$.}\label{fig3}
\end{figure}

In the previous subsection, we can see that even the SPPs can be trapped by the barrier, it can not propagate stably in the barrier, i.e., the stable propagation modes of the barrier can not be excited by inputting a single nonlinear SPPs on one side of the barrier. However, when inputting two nonlinear SPPs symmetrically on the two boundaries, as shown in Fig.~\ref{fig3}(a), the initial condition of the nonlinear SPPs are $v={\rm sech}( \xi -12)e^{-i0.5 \xi}+{\rm sech}( \xi +12)e^{i0.5 \xi }$, with $V_0=5$ and $l=4$, we can see that, after the two nonlinear SPPs incident on the two boundaries of the barrier, partial energy reflects out of the system, but the parts transit into the barrier can produce a periodic intensity distribution in transverse direction $\xi$, i.e., a 'standing wave' shape mode can be excited inside the barrier, and it can propagate stably for a long distance. The blue solid line shown in Fig.~\ref{fig3}(b) is the profile of intensity $|v|^2$ as a function of $\xi$ at $s=30$, the red solid line on top of the figure is the corresponding barrier, and the black dashed lines denote the boundary of the barrier. Such a 'standing wave'-like SPPs can be used to design phase or intensity SPPs gratings, which have great application potentials in SPPs based precision measurement at micro/nano scale.

The physical reason for this stable propagation and periodic distributed mode is that, inside the barrier, the two SPPs are reflected back and forth, respectively, thus, they can meet periodically, and lead to quantum interference, such interference can generate a periodic intensity distribution.

However, our in-depth study find that, not all the periodic modes excited in the barrier can propagate stably, it depends on three major factors, which are nonlinearity, half width of the barrier and phase difference of the initial nonlinear SPPs, respectively.

\begin{figure}
\includegraphics[width=0.45\textwidth]{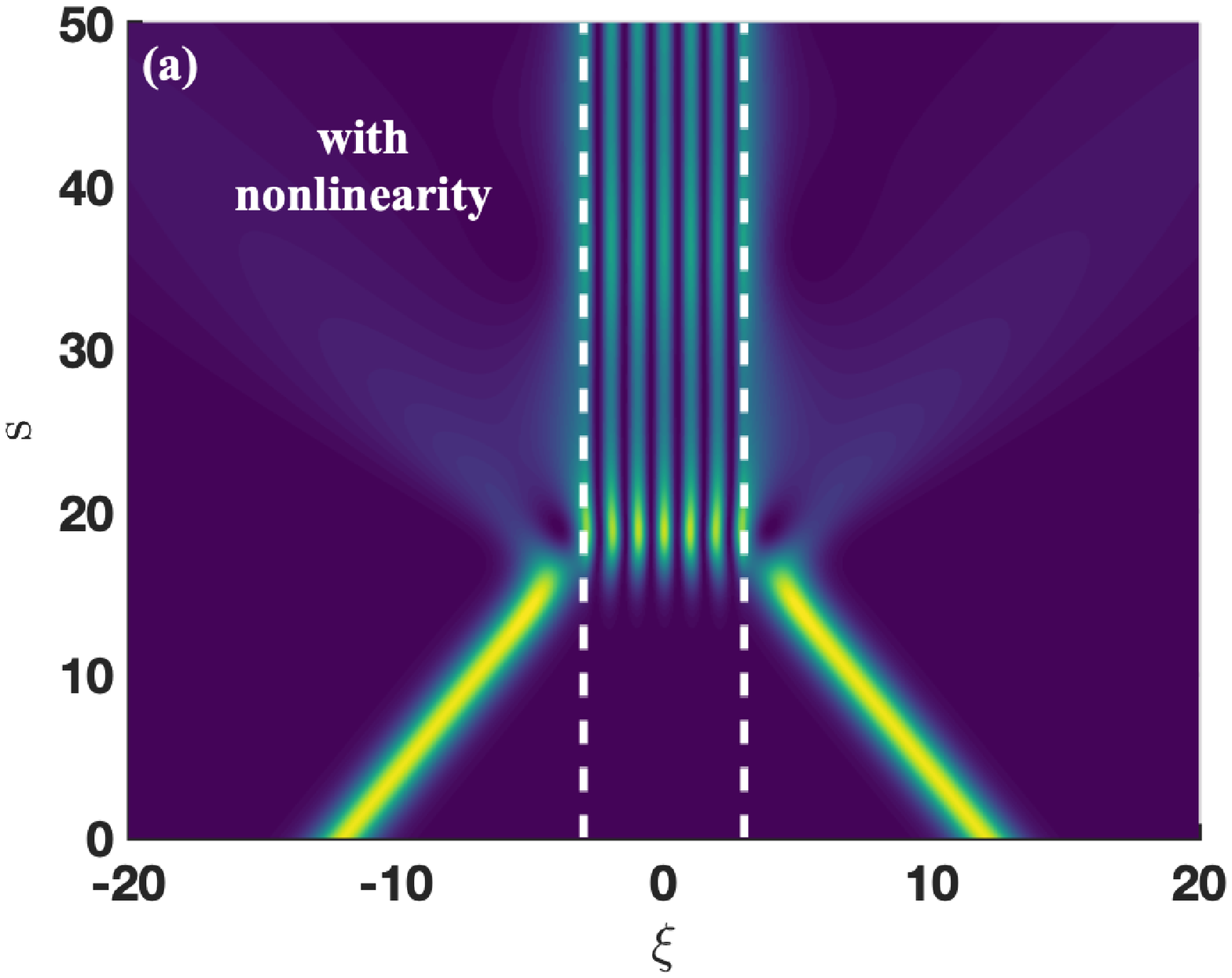}
\includegraphics[width=0.45\textwidth]{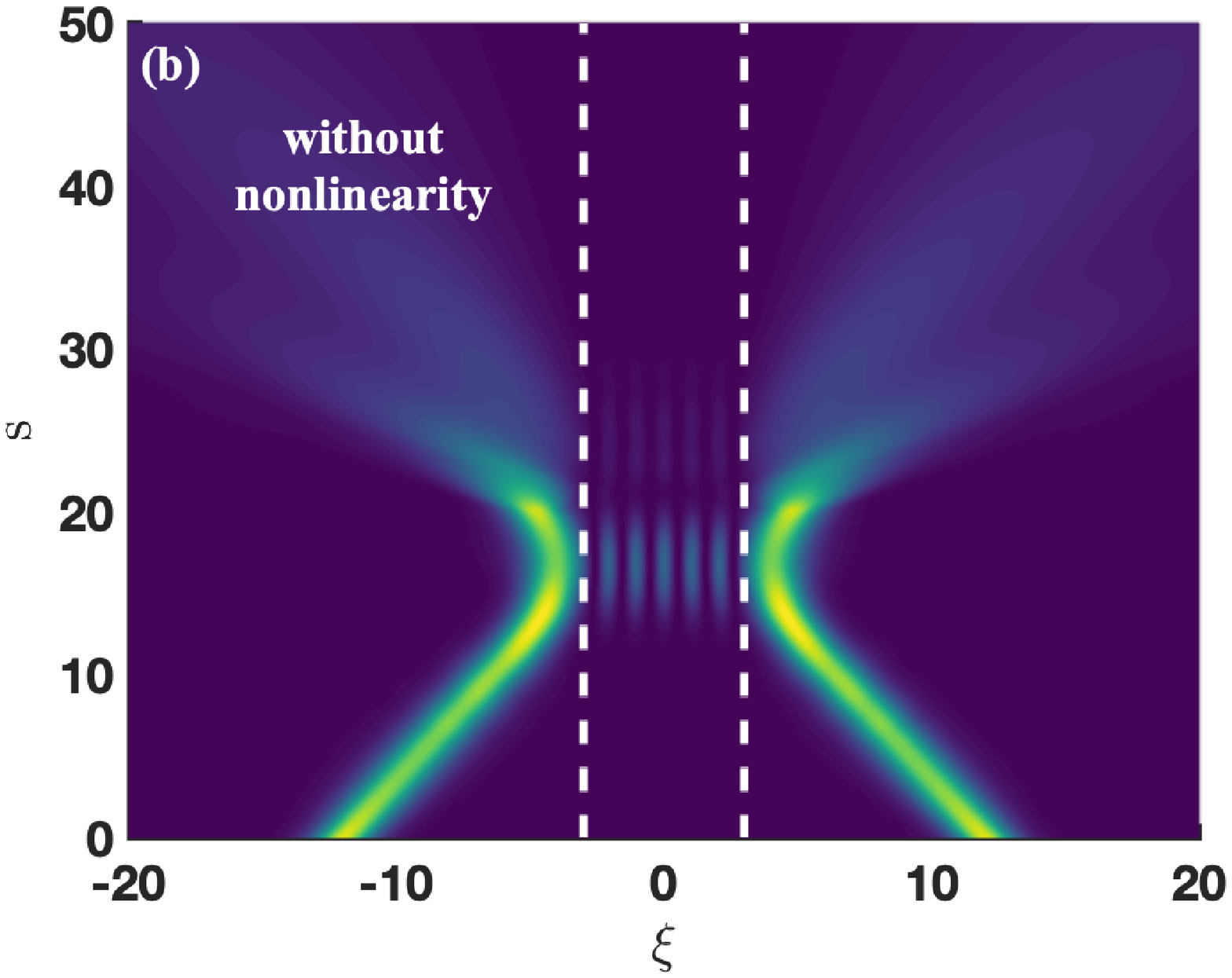}
\includegraphics[width=0.45\textwidth]{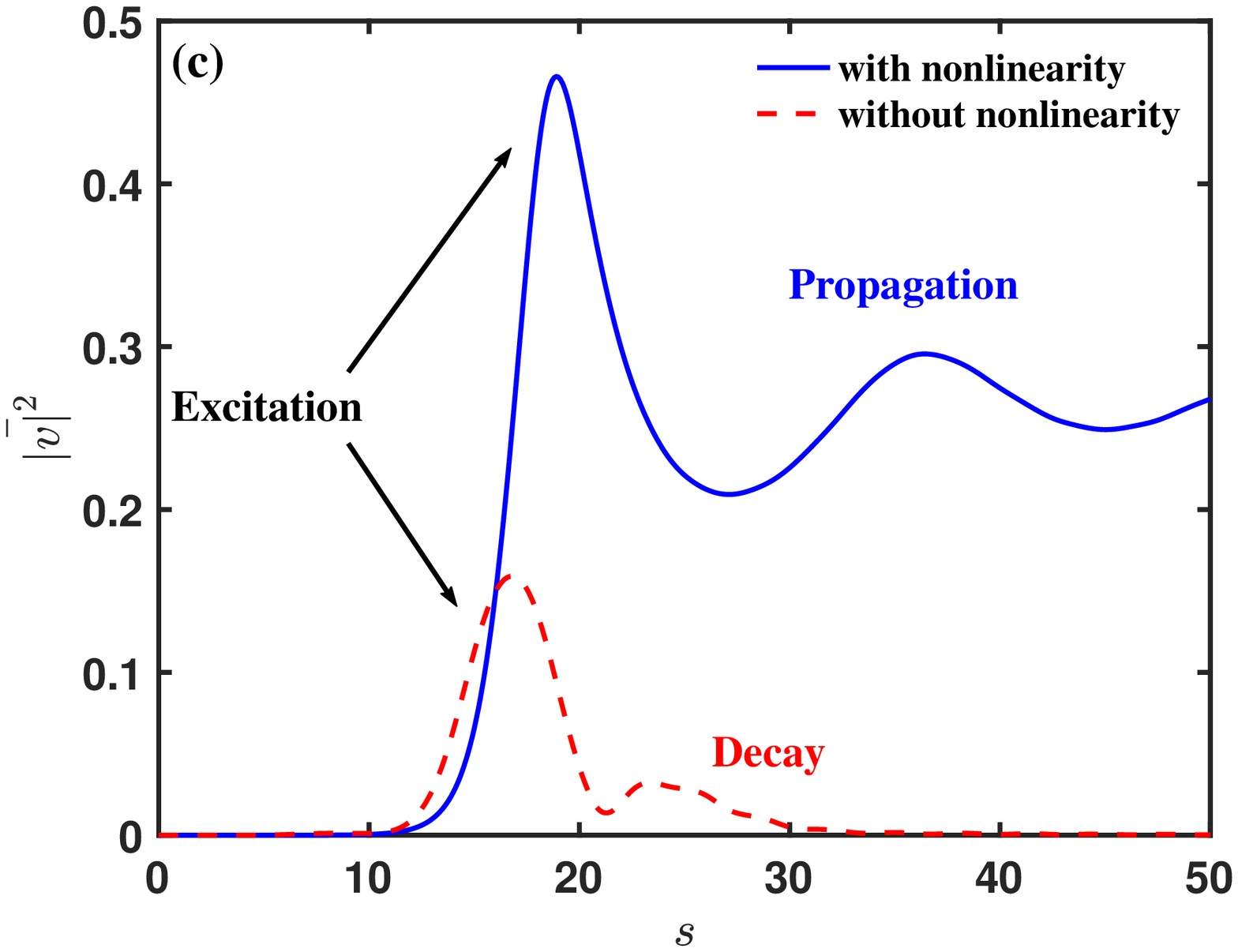}
\caption{(a) and (b) express the detail propagation circumstance of the propagation modes in the potential barrier for the case with the nonlinearity modulation or without it, respectively. (c) The variation tendency about the average intensity of the propagation modes with the growth of propagation distance $s$ under the circumstance of nonlinearity modulation or without it.}
\label{fig4}
\end{figure}

Shown in Fig.~\ref{fig4}(a) and (b) are intensity $|v|^2$ as functions of $s$ and $\xi$ with and without nonlinearity, respectively. The initial conditions are the same as that used in Fig.~\ref{fig3} but with $V_0=5$ and $l=3$. We can see that, when the two nonlinear SPPs enter the barrier and interact around the position at $s=20$, the periodic modes can both be excited, for the case with nonlinearity (Fig.~\ref{fig4}(a)), intensity of the mode out of the barrier at $s=50$ is much larger than that in the case without nonlinearity (Fig.~\ref{fig4}(b)). To be more intuitive, we define the average intensity of the mode inside the barrier in $\xi$ direction as
\begin{equation}\label{vaver}
  \bar{|v|^2}(s)=\frac{\int_{-l}^{l}|v(s,\xi)|^2d\xi}{\int_{-l}^{l}d\xi}.
\end{equation}

In Fig.~\ref{fig4}(c), the blue solid (red dashed) line is $\bar{|v|^2}$ as a function of $s$ for the case with nonlinearity (without nonlinearity). The first peaks from left denote the position where the modes be excited for the two cases, but after the excitation process, we can see that the blue solid line first drops, then oscillates and flattens as increase of $s$ due to the modulation of the nonlinearity, which means the periodic mode be excited can propagate stably. However, when the nonlinearity of the system is absence, the average intensity $\bar{|v|^2}$ keeps decaying to a quite small value with the increase of $s$, shown as red dashed line after $s=20$, which means the periodic mode even can be excited in the system without nonlinearity, it can not propagate stably in the barrier, which can be treated as decay modes, and the parts of the nonlinear SPPs out of the barrier are broadening due to the diffraction effect.

\begin{figure}[H]
  \centering
  \includegraphics[width=1\textwidth]{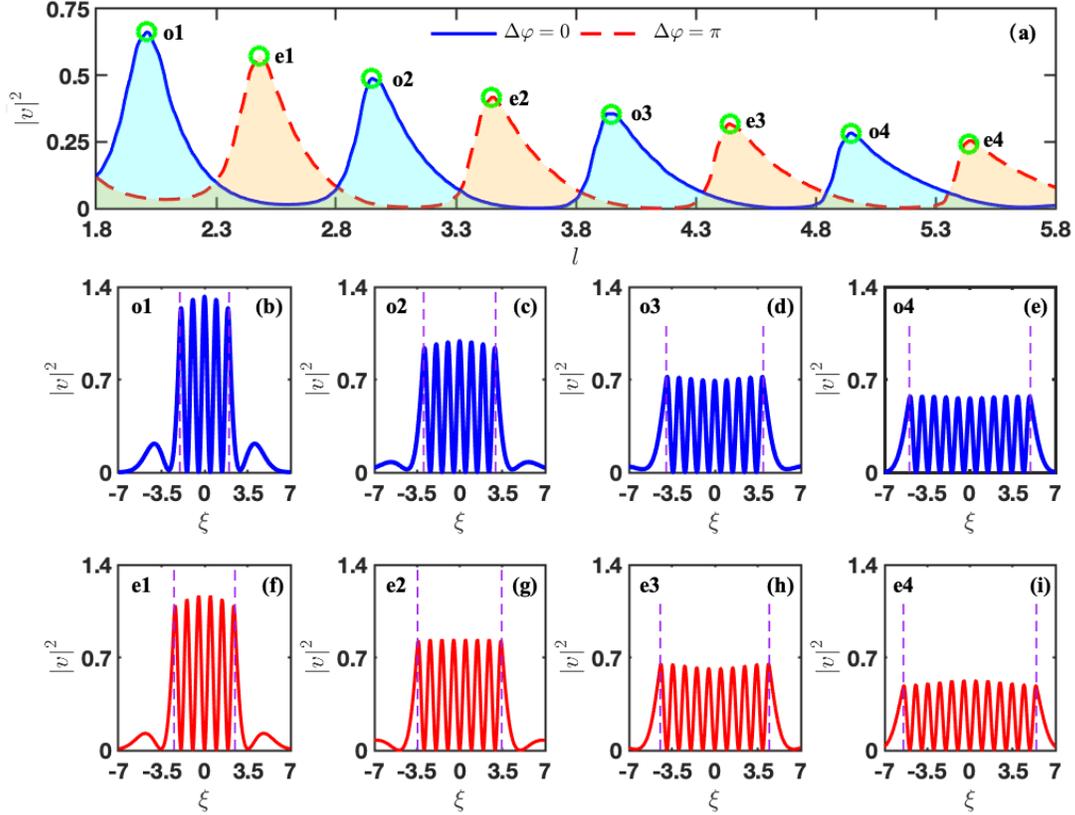}\\
  \caption{(a) The changing relations between the average intensity $\bar{|v|^2}$ and the half width of the barrier potential $l$ for the case with different phase difference. (b) - (i) show the corresponding profile of $|v|^2$ with $\Delta \varphi=0$ (blue solid line) and $\Delta \varphi=\pi$ (red solid line), respectively.}
\label{fig5}
\end{figure}

In addition to the effect of nonlinearity, we study the influence of width of the barrier to the propagation modes in the next. Shown in Fig.~\ref{fig5}(a) are the average intensity $\bar{|v|^2}$ at $s=20$ as functions of half width of the barrier $l$, in which, we have selected two typical phase difference $\Delta \varphi=0$ (blue solid line, attractive interaction of nonlinear SPPs inside the barrier) and $\Delta \varphi=\pi$ (red dashed line, repulsive interaction of nonlinear SPPs inside the barrier),
and the initial condition of the two nonlinear SPPs are the same as that given in Fig.~\ref{fig3}(a), the height of the barrier is $V_0=5$ for both cases. From blue solid line in Fig.~\ref{fig5}(a), we can find that the average intensity $\bar{|v|^2}$ has quasi periodic peaks with the increase of $l$, and the value of the peaks continuously decrease with the increase of $l$, which means with the increase of $l$, quasi periodic coherence enhancement occurs in the barrier. The peaks mean that propagation modes can be excited in the barrier with corresponding half widths, but the dips in the blue solid line mean decay modes are excited, i.e., even modes can be excited, they can not propagate for a long distance. In addition, we can obtain the similar results for $\Delta \varphi=\pi$, shown as red dashed line in Fig.~\ref{fig5}(a).

Shown in Fig.~\ref{fig5}(b)-(e) are the wave shapes $|v|^2$ as functions of $\xi$ at $s=20$, the corresponding half width $l$ are selected from positions of the first four peaks of the blue solid line in Fig.~\ref{fig5}(a), the purple dash lines denote the boundaries of the barriers. We can see that there are three complete peaks inside the barrier in Fig.~\ref{fig5}(b), and the number of complete peaks are 5, 7, 9 in Fig.~\ref{fig5}(c), (d) (e), respectively, we define those modes as odd modes. For $\Delta \varphi=0$, we can find that the number of complete peaks are odd numbers. In contrary, the number of complete peaks are even numbers for $\Delta \varphi=\pi$, as shown in Fig.~\ref{fig5}(f)-(i). Thus, by adjusting the half width $l$ and phase difference $\Delta \varphi$, we can modulate structure of the propagation modes in the barrier.

In addition, it is worth noting that we have selected two typical phase differences in the simulation of last figure, and peaks for $\Delta \varphi=0$ approximately correspond to dips for $\Delta \varphi=\pi$ and vice versa. Such a character can be used to design an optical switch or XNOR logic gate based on phase difference. Next, we will study the average intensity $\bar {|v|^2}$ when the phase difference $\Delta \varphi$ changes continuously, and give the theoretical scheme to realize the optical switch.

\begin{figure}
\includegraphics[width=0.45\textwidth]{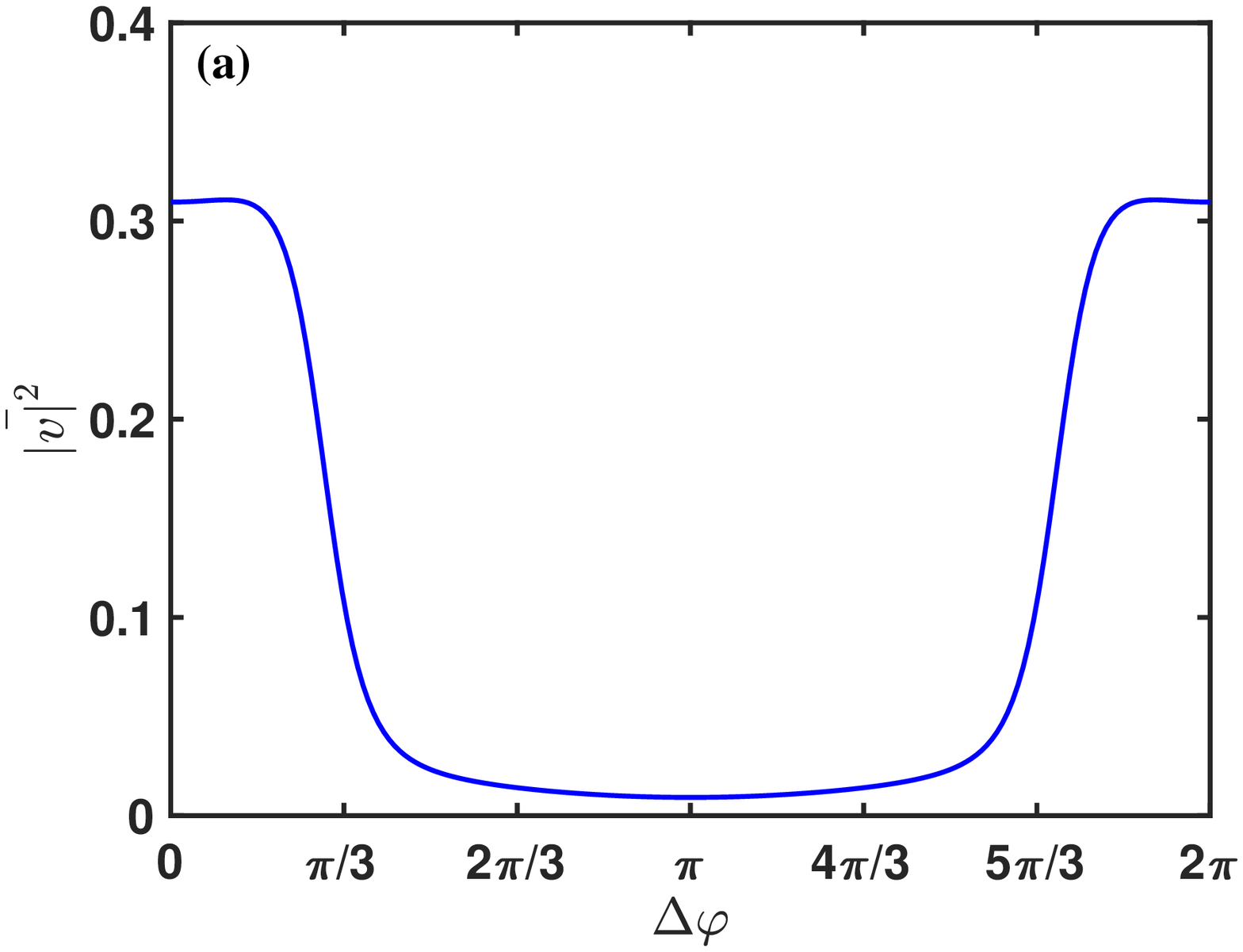}
\includegraphics[width=0.45\textwidth]{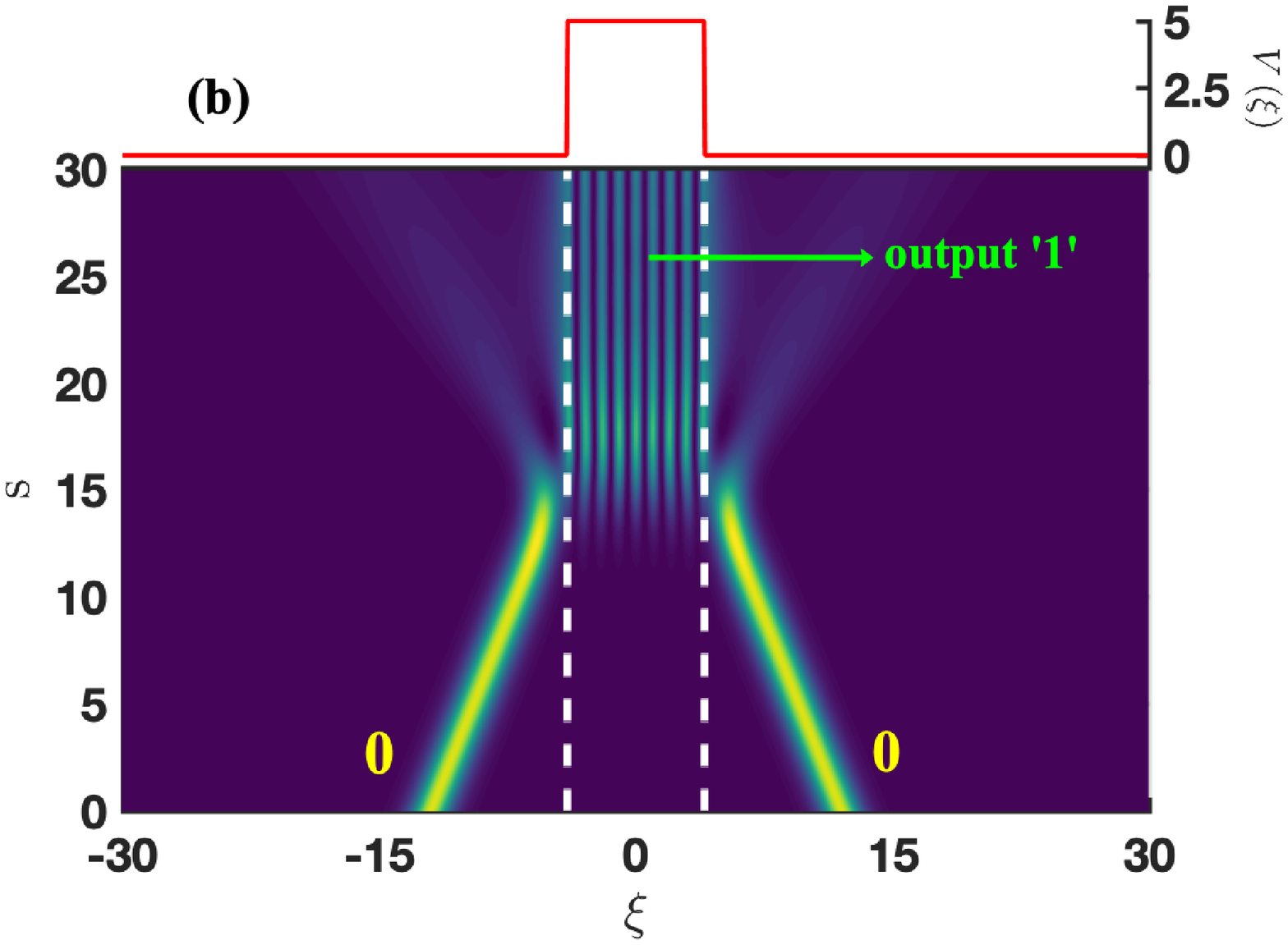}
\includegraphics[width=0.45\textwidth]{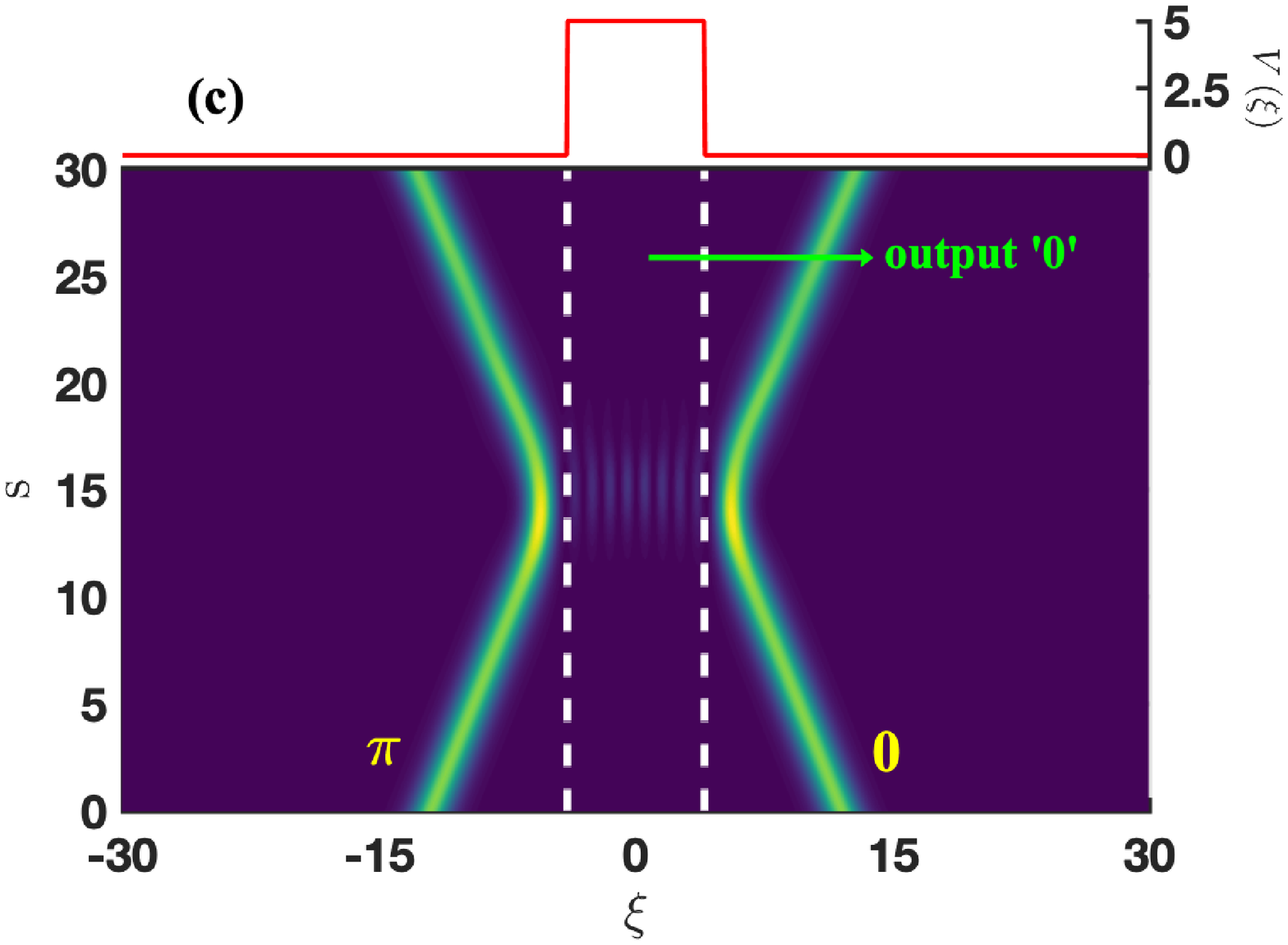}
\caption{(a) The average intensity $\bar {|v|^2}$ as a function of the phase difference $\Delta \varphi$ at {$s=20$}. (b) and (c) are the intensity $|v|^2$ as functions of $s$ and $\xi$, which describe the excitation of nonlinear SPPs in the rectangular barrier.}
\label{fig6}
\end{figure}

Shown in Fig.~\ref{fig6}(a) is the average intensity $\bar {|v|^2}$ as a function of the phase difference $\Delta \varphi$ at $s=20$. Initial condition of the nonlinear SPPs is $v={\rm sech}( \xi -12) e^{-i0.5 \xi}+{\rm sech}( \xi +12) e^{i0.5 \xi+ \Delta \varphi}$, and the height and half width are $V_0=5$ and $l=4$, respectively. With the phase difference increase, we can see that the average intensity $\bar {|v|^2}$ is large at first, and changes slowly, then it drops rapidly to a small value (close to 0) in the region $[\pi/6,\pi/2]$, and then keeps changing slowly, rises sharply in the region $[3\pi/2,11\pi/6]$. This step change curve can be used to design an optical switch.

Shown in Fig.~\ref{fig6}(b) and (c) are intensity $|v|^2$ as functions of $s$ and $\xi$, parameters are the same as that used in Fig.~\ref{fig6}(a). In Fig.~\ref{fig6}(b), we can see that a large partial energy of the nonlinear SPPs couples into the barrier when the phase difference $\Delta \varphi=0$, and the output strength in the end of the barrier is quite large, and we denote this state as output '1'. However, when the phase difference $\Delta \varphi=\pi$, as shown in Fig.~\ref{fig6}(c), most energy of the nonlinear SPPs reflects out of the system, and the output strength in the end of the barrier is approach to 0, and we denote this state as output '0'. Thus, a phase depended optical switch is obtained.

\section{Conclusion}\label{Conclusion}
In summary, we have proposed a scheme to study the propagation and excitation properties of nonlinear SPPs under modulation of the rectangular potential barrier. Based on multi scale method, we have derived the NLSE in the system, and the Stark field applied into the system produced an equivalent rectangular barrier to affect the propagation properties of the nonlinear SPPs. When incident a single nonlinear SPPs, we have realized transmission, trapping and reflection of the nonlinear SPPs. When two nonlinear SPPs symmetrically inputting the system, we found that a periodic mode can be excited in the rectangular barrier, and excitation of such modes are depending on nonlinearity in the system, half width of the barrier and phase difference of the initial nonlinear SPPs. Based on the related properties, we have designed an optical switch based on phase difference of the nonlinear SPPs. These results not only provide a theoretical basis for the study of the interaction between nonlinear SPPs and external potentials, but also have broad application prospects in the field of optical information at micro/nano scale.

\section*{Acknowledgments}
This work was supported by Natural Science Foundation of Shandong Province (Grant No. ZR2021MA035). Xiangchun Tian and Yundong Zhang contribute to this work equally.

\section*{Appendix}
\appendix
\section{THE TM MODE IN THE WAVEGUIDE SYSTEM}\label{Appendix A}
The quantized probe field can be expressed as~\cite{PhysRevLett.101.263601,PhysRevA.91.023803}

\begin{equation}
{{\bf{E}}}\left( {{\bf{r}},t} \right) = \left\{ {\begin{array}{*{20}{l}}
{\left( {k{{\bf{e}}_x} - i{k_1}{{\bf{e}}_z}} \right)\frac{c}{{{\varepsilon _1}{\omega _l}}}\sqrt {\frac{{\hbar {\omega _l}}}{{{\varepsilon _0}{L_x}{L_y}{L_z}}}} \hat a\left( {{\omega _p}} \right){e^{ - {k_1}x + i\left( {kz - {\omega _l}t} \right)}} + c.c.}&{z > 0}\\
{\left( {k{{\bf{e}}_x} + i{k_2}{{\bf{e}}_z}} \right)\frac{c}{{{\varepsilon _2}{\omega _l}}}\sqrt {\frac{{\hbar {\omega _l}}}{{{\varepsilon _0}{L_x}{L_y}{L_z}}}} \hat a\left( {{\omega _p}} \right){e^{{k_2}x + i\left( {kz - {\omega _l}t} \right)}} + c.c.}&{z < 0.}
\end{array}} \right.
\end{equation}

Here the expression of the wave vector is $k_j^2 = k^2 - \omega _l^2{\varepsilon _j}{\mu _j}/{c^2}$( $j = 1$ for the hot atomic gas and $j=2$ for the NIMM) satisfies the boundary condition ${\varepsilon _2}{k_1} =  - {\varepsilon _1}{k_2}$ at $x=0$. The propagation constant of probe field can be expressed as ${k}(\omega _l) = \left( {{\omega _l}/c} \right){\left[ {{\varepsilon _1}{\varepsilon _2}\left( {{\varepsilon _1}{\mu _2} - {\varepsilon _2}{\mu _1}} \right)/\left( {\varepsilon _1^2 - \varepsilon _2^2} \right)} \right]^{1/2}}$. $\hat a\left( {{\omega _p}} \right)$ is the creation operator of TM photons. $L_y$ and $L_z$ are the lengths of the NIMM-hot atomic gas interface in the $y$ and $z$ directions, respectively. $L_x$ is the effective mode length characterizing the electromagnetic field confinement in the $x$ direction which is defined as

\begin{equation}
{L_x} \equiv \sum\limits_{j = 1,2} {\frac{1}{{2{\mathop{\rm Re}\nolimits} \left( {{k_j}} \right)}}} \left[ {\frac{{{{\tilde \varepsilon }_j}}}{{{{\left| {{\varepsilon _j}} \right|}^2}}}\frac{{{c^2}}}{{\omega _l^2}}\left( {{{\left| k \right|}^2} + {{\left| {{k_j}} \right|}^2}} \right) + {{\tilde \mu }_j}} \right] ,
\end{equation}
with $\tilde \varepsilon  \equiv {\mathop{\rm Re}\nolimits} \left[ {\partial \left( {{\omega _l}\varepsilon } \right)/\partial {\omega _l}} \right]$, $\tilde \mu  \equiv {\mathop{\rm Re}\nolimits} \left[ {\partial \left( {{\omega _l}\mu } \right)/\partial {\omega _l}} \right]$.

\section{THE BLOCH EQUATIONS IN THE INTERACTION PICTURE}\label{Appendix B}
The explicit expressions of the Bloch equations in the interaction picture can be expressed as follows:
\begin{subequations}\label{density matrix equations}
\begin{align}
&{i\frac{\partial }{{\partial t}}{\sigma _{11}} - i{\Gamma _{13}}{\sigma _{33}} + {e^{-i\theta_p^*}} {\zeta ^*}\left( x \right)\Omega _p^*{\sigma _{31}} - {e^{i\theta_p}} \zeta \left( x \right){\Omega _p}\sigma _{31}^* = 0},\\
&{i\frac{\partial }{{\partial t}}{\sigma _{22}} - i{\Gamma _{23}}{\sigma _{33}} + \Omega _c^*{\sigma _{32}} - {\Omega _c}\sigma _{32}^* = 0},\\
&{i\frac{\partial }{{\partial t}}{\sigma _{33}} + i {{\Gamma _{3}}} {\sigma _{33}} - {e^{-i\theta_p^*}} {\zeta ^*}\left( x \right)\Omega _p^*{\sigma _{31}} + {e^{i\theta_p}} \zeta \left( x \right){\Omega _p}\sigma _{31}^* - \Omega _c^*{\sigma _{32}} + {\Omega _c}\sigma _{32}^* = 0},\\
&{\left( {i\frac{\partial }{{\partial t}} + {d_{21}}} \right){\sigma _{21}} + \Omega _c^*{\sigma _{31}} - {e^{i\theta_p}} \zeta \left( x \right){\Omega _p}\sigma _{32}^* = 0},\\
&{\left( {i\frac{\partial }{{\partial t}} + {d_{31}}} \right){\sigma _{31}} + {e^{i\theta_p}} \zeta \left( x \right){\Omega _p}\left( {{\sigma _{11}} - {\sigma _{33}}} \right) + {\Omega _c}{\sigma _{21}} = 0},\\
&{\left( {i\frac{\partial }{{\partial t}} + {d_{32}}} \right){\sigma _{32}} + {\Omega _c}\left( {{\sigma _{22}} - {\sigma _{33}}} \right) + {e^{i\theta_p}} \zeta \left( x \right){\Omega _p}\sigma _{21}^* = 0},
\end{align}
\end{subequations}
where ${d_{21}} =  - \left( {{\mathbf{k}_p} - {\mathbf{k}_c}} \right) \cdot \mathbf{v} + {{\Delta }_2} - {{\Delta }_1} + i{\gamma _{21}}+\frac{1}{2}{\alpha _{21}}{\left| {{E_s}} \right|^2}, ~ {d_{31}} =  - {\mathbf{k}_p} \cdot \mathbf{v} + {{\Delta }_3} - {{\Delta }_1} + i{\gamma _{31}}+\frac{1}{2}{\alpha _{31}}{\left| {{E_s}} \right|^2}, ~ {d_{32}} =  - {\mathbf{k}_c} \cdot \mathbf{v} + {{\Delta }_3} - {{\Delta }_2} + i{\gamma _{32}}+\frac{1}{2}{\alpha _{32}}{\left| {{E_s}} \right|^2}$, with ${\Gamma _j} = \sum\nolimits_{E_l < E_j} { {\Gamma _{lj}}}, ~{\gamma _{jl}} = \left( {{\Gamma _j} + {\Gamma _l}} \right)/2$. Here ${{\Gamma _j}}$ is the total spontaneous emission decay rate and ${\gamma _{jl}}$ is the atomic decay rate.

\section{EXPRESSIONS OF EACH ORDER SOLUTIONS}\label{Appendix C}

\subsection{First-order solutions}
The first-order solutions reads
\begin{subequations}
\begin{align}
&\Omega _p^{\left( 1 \right)} = {F}{e^{i{\theta }}},\\
&\sigma _{21}^{\left( 1 \right)} =  - \frac{{\Omega _c^*}}{{{D}}}{F} {{e^{i{(\theta + \theta_p)}}}} \zeta \left( x \right),\\
&\sigma _{31}^{\left( 1 \right)} = \frac{{\left( {\omega  + d_{21}^{\left( 0 \right)}} \right)}}{{{D}}} {F} {{e^{i{(\theta + \theta_p)}}}} \zeta \left( x \right),
\end{align}
\end{subequations}
where ${\theta } = {K}\left( \omega  \right){z_0} - \omega {t_0},{{D} = {{\left| {{\Omega _c}} \right|}^2} - \left( {\omega  + d_{21}^{\left( 0 \right)}} \right)\left( {\omega  + d_{31}^{\left( 0 \right)}} \right)}$, with all other $\sigma _{ij}^{(1)}$ being zero. Here $F$ is the envelope function of the slow variables $y_1,z_2$ and $t_2$. The expression of the linear dispersion relation $K(\omega)$ is
\begin{equation}
{K}\left( \omega  \right) = \frac{\omega }{c}\frac{1}{{{n_{\rm eff}}}} + {\kappa _{13}}\int_{ - \infty }^\infty  {d(kv)f\left( kv \right)\left\langle {\frac{{\left( {\omega  + d_{21}^{\left( 0 \right)}} \right)}}{{{D}}}\zeta \left( x \right)} \right\rangle }  .
\end{equation}

\subsection{Second-order solutions}
The expressions of the second-order solutions are given by
\begin{subequations}
\begin{align}
&{\sigma _{11}^{\left( 2 \right)} = a_{11}^{\left( 2 \right)}{{\left| {\zeta \left( x \right)} \right|}^2}{{\left| {{F}} \right|}^2}{e^{ - 2{{\tilde \alpha }}{z_2}}}},\\
&{\sigma _{22}^{\left( 2 \right)} = a_{22}^{\left( 2 \right)}{{\left| {\zeta \left( x \right)} \right|}^2}{{\left| {{F}} \right|}^2}{e^{ - 2{{\tilde \alpha }}{z_2}}}},\\
&{\sigma _{32}^{\left( 2 \right)} = a_{32}^{\left( 2 \right)}{{\left| {\zeta \left( x \right)} \right|}^2}{{\left| {{F}} \right|}^2}{e^{ - 2{{\tilde \alpha }}{z_2}}}},\\
&{\sigma _{33}^{\left( 2 \right)} = a_{33}^{\left( 2 \right)}{{\left| {\zeta \left( x \right)} \right|}^2}{{\left| {{F}} \right|}^2}{e^{ - 2{{\tilde \alpha }}{z_2}}}},\\
\end{align}
\end{subequations}
where ${{\tilde \alpha }} = {\epsilon ^{ - 2}}{\mathop{\rm Im}\nolimits} \left[ {{{K}\left( \omega  \right)+k(\omega_p)}} \right]$ and
\begin{subequations}
\begin{align}
&a_{11}^{\left( 2 \right)} =  - a_{22}^{\left( {\rm{2}} \right)} - a_{33}^{\left( 2 \right)},\\
&a_{22}^{\left( 2 \right)} = \frac{{\frac{{i{\Gamma _{23}}}}{{{{\left| {{\Omega _c}} \right|}^2}}}a_{33}^{\left( 2 \right)} + \frac{1}{{\omega  + d_{32}^{*\left( 0 \right)}}}\left[ {a_{33}^{\left( 2 \right)} + \frac{1}{{{D}}}} \right] - \frac{1}{{\omega  + d_{32}^{\left( 0 \right)}}}\left[ {a_{33}^{\left( 2 \right)} + \frac{1}{{D^*}}} \right]}}{{\frac{1}{{\omega  + d_{32}^{*\left( 0 \right)}}} - \frac{1}{{\omega  + d_{32}^{\left( 0 \right)}}}}},\\
&a_{32}^{\left( 2 \right)} = \frac{{{\Omega _c}}}{{\omega  + d_{32}^{\left( 0 \right)}}}\left[ {a_{33}^{\left( 2 \right)} - a_{22}^{\left( 2 \right)} + \frac{1}{{D^*}}} \right],\\
&a_{33}^{\left( 2 \right)} = \frac{1}{{i{\Gamma _{13}}}}\left[ {\frac{{\omega {\rm{ + }}d_{21}^{\left( 0 \right)}}}{{{D}}} - \frac{{\omega  + d_{21}^{*\left( 0 \right)}}}{{D^*}}} \right].
\end{align}
\end{subequations}

\subsection{Third-order solutions}
In the third-order approximation, we obtain the explicit expression of $W_1$ and $W_2$ in Eq.~(\ref{nls}), which read
\begin{subequations}
\begin{align}
&{{W_1} = {\kappa _{13}}\int_{ - \infty }^\infty  {d(kv)f\left( kv \right)\left\langle {\frac{{\left( {\omega  + d_{21}^{\left( 0 \right)}} \right)\left( {a_{11}^{\left( 2 \right)} - a_{33}^{\left( 2 \right)}} \right) + {\Omega _c}a_{32}^{*\left( 2 \right)}}}{{{D}}}\zeta \left( x \right){{\left| {\zeta \left( x \right)} \right|}^2}} \right\rangle } },\\
&{{W_2} = {\kappa _{13}}\int_{ - \infty }^\infty  {d(kv)f\left( kv \right)\left\langle {\frac{{{{\left( {\omega  + d_{21}^{\left( 0 \right)}} \right)}^2}{\alpha _{31}} + {{\left| {{\Omega _c}} \right|}^2}{\alpha _{21}}}}{{2D^2}}\zeta \left( x \right)} \right\rangle } } .
\end{align}
\end{subequations}
%


\end{document}